\documentclass[lettersize,journal]{IEEEtran}
\usepackage{amsmath,amsfonts}
\usepackage{algorithmic}
\usepackage{algorithm}
\usepackage{array}
\usepackage{textcomp}
\usepackage{stfloats}
\usepackage{url}
\usepackage{verbatim}
\usepackage{graphicx}
\usepackage{cite}
\usepackage{colortbl}
\usepackage{makecell}
\usepackage{booktabs}
\usepackage{multirow}
\usepackage{pifont}

\usepackage{arydshln}
\usepackage{subcaption}
\usepackage{rotating}
\usepackage{amsmath,amsfonts}
\usepackage{algorithmic}
\usepackage{algorithm}
\usepackage{array}
\usepackage{amsmath,amsfonts}
\usepackage{algorithmic}
\usepackage{algorithm}
\usepackage{array}
\usepackage{multicol}
\usepackage{multirow}
\usepackage{textcomp}
\usepackage{stfloats}
\usepackage{verbatim}
\usepackage{graphicx}
\usepackage{cite}
\usepackage{color}
\usepackage{graphicx}
\usepackage{amsmath}
\usepackage{amssymb}
\usepackage{booktabs}
\usepackage{stfloats}
\usepackage{soul}
\usepackage{graphicx}
\usepackage{amsmath}
\usepackage{amssymb}
\usepackage{booktabs}
\usepackage{colortbl}
\usepackage{times}
\usepackage{epsfig}
\usepackage{multirow} 
\usepackage{overpic}
\usepackage{hhline}
\usepackage{xcolor}
\usepackage{arydshln} 
\usepackage{pifont}
\usepackage{enumitem}
\usepackage{colortbl}
\usepackage{makecell}
\usepackage{verbatim}
\usepackage{bm}
\usepackage{caption,subcaption}
\usepackage{bbding}
\usepackage{multirow}

\usepackage{amsmath,amsfonts}
\usepackage{algorithmic}
\usepackage{algorithm}
\usepackage{array}
\usepackage{multicol}
\usepackage{multirow}
\usepackage{amsmath,amsfonts}
\usepackage{algorithmic}
\usepackage{algorithm}
\usepackage{array}
\usepackage{textcomp}
\usepackage{stfloats}
\usepackage{url}
\usepackage{verbatim}
\usepackage{graphicx}
\usepackage{cite}
\usepackage{colortbl}
\usepackage{makecell}
\usepackage{booktabs}
\usepackage{multirow}
\usepackage{pifont}
\usepackage[hidelinks,colorlinks,
urlcolor=black,
            linkcolor=black,
            anchorcolor=black,
            citecolor=blue]{hyperref}
\usepackage{arydshln}
\usepackage{subcaption}
\usepackage{rotating}
\usepackage{xcolor}
\usepackage{color}

\usepackage{stfloats}
\usepackage{url}
\usepackage{verbatim}
\usepackage{graphicx}
\usepackage{cite}
\usepackage{graphicx}
\usepackage{amsmath}
\usepackage{amssymb}
\usepackage{booktabs}
\usepackage{stfloats}
\usepackage{graphicx}
\usepackage{amsmath}
\usepackage{amssymb}
\usepackage{booktabs}

\begin{document}

\title{Efficient Real-world Image Super-Resolution Via Adaptive Directional Gradient Convolution}

\author{
Long Peng$^{1}$,
Yang Cao$^{*1}$,
Renjing Pei$^{2}$,
Wenbo Li$^{2}$,
Jiaming Guo$^{2}$,
Xueyang Fu$^{1}$,
Yang Wang†$^{1}$,
Zheng-Jun Zha$^{1}$
{ 1. University of Science and Technology of China, China. } \\
{ 2. Huawei Noah’s Ark Lab.}
\thanks{* Long Peng and Yang Cao contributed equally to this paper. † Yang Wang is the corresponding author of this paper. Long Peng, Yang Wang, Yang Cao, Xueyang Fu, and Zheng-Jun Zha are with the University of Science and Technology of China, Hefei 230026, China. (e-mail: longp2001@mail.ustc.edu.cn, ywang120@ustc.edu.cn, forrest@ustc.edu.cn, zhazj@ustc.edu.cn). Renjing Pei, Wenbo Li, and Jiaming Guo are with Huawei Noah's Ark Lab (e-mail: peirenjing@huawei.com, liwenbo50@huawei.com, and guojiaming5@huawei.com).}

\thanks{This work was supported by the National Natural Science Foundation of China (NSFC) under Grants 62206262.}

}


\markboth{Journal of \LaTeX\ Class Files,~Vol.~14, No.~8, August~2021}%
{Shell \MakeLowercase{\textit{et al.}}: A Sample Article Using IEEEtran.cls for IEEE Journals}

\maketitle

\begin{abstract}
Real-world image super-resolution (Real-SR) endeavors to produce high-resolution images with rich details while mitigating the impact of multiple degradation factors. Although existing methods have achieved impressive achievements in detail recovery, they still fall short when addressing regions with complex gradient arrangements due to the intensity-based linear weighting feature extraction manner. Moreover, the stochastic artifacts introduced by degradation cues during the imaging process in real LR increase the disorder of the overall image details, further complicating the perception of intrinsic gradient arrangement. To address these challenges, we innovatively introduce kernel-wise differential operations within the convolutional kernel and develop several learnable directional gradient convolutions. These convolutions are integrated in parallel with a novel linear weighting mechanism to form an Adaptive Directional Gradient Convolution (DGConv), which adaptively weights and fuses the basic directional gradients to improve the gradient arrangement perception capability for both regular and irregular textures. Coupled with DGConv, we further devise a novel equivalent parameter fusion method for DGConv that maintains its rich representational capabilities while keeping computational costs consistent with a single Vanilla Convolution (VConv), enabling DGConv to improve the performance of existing super-resolution networks without incurring additional computational expenses. To better leverage the superiority of DGConv, we further develop an Adaptive Information Interaction Block (AIIBlock) to adeptly balance the enhancement of texture and contrast while meticulously investigating the interdependencies, culminating in the creation of a Directional Gradient Perceiving Network (DGPNet) for Real-SR through simple stacking. Comparative results with 15 state-of-the-art (SOTA) super-resolution methods across three public datasets underscore the effectiveness and efficiency of our proposed approach. The source code will be made publicly available at \href{https://github.com/peylnog/ERealSR-DGPNet}{https://github.com/peylnog/ERealSR-DGPNet}.
\end{abstract}


\section{Introduction}

\begin{figure}[tbp]
    \includegraphics[width=1.0\linewidth]{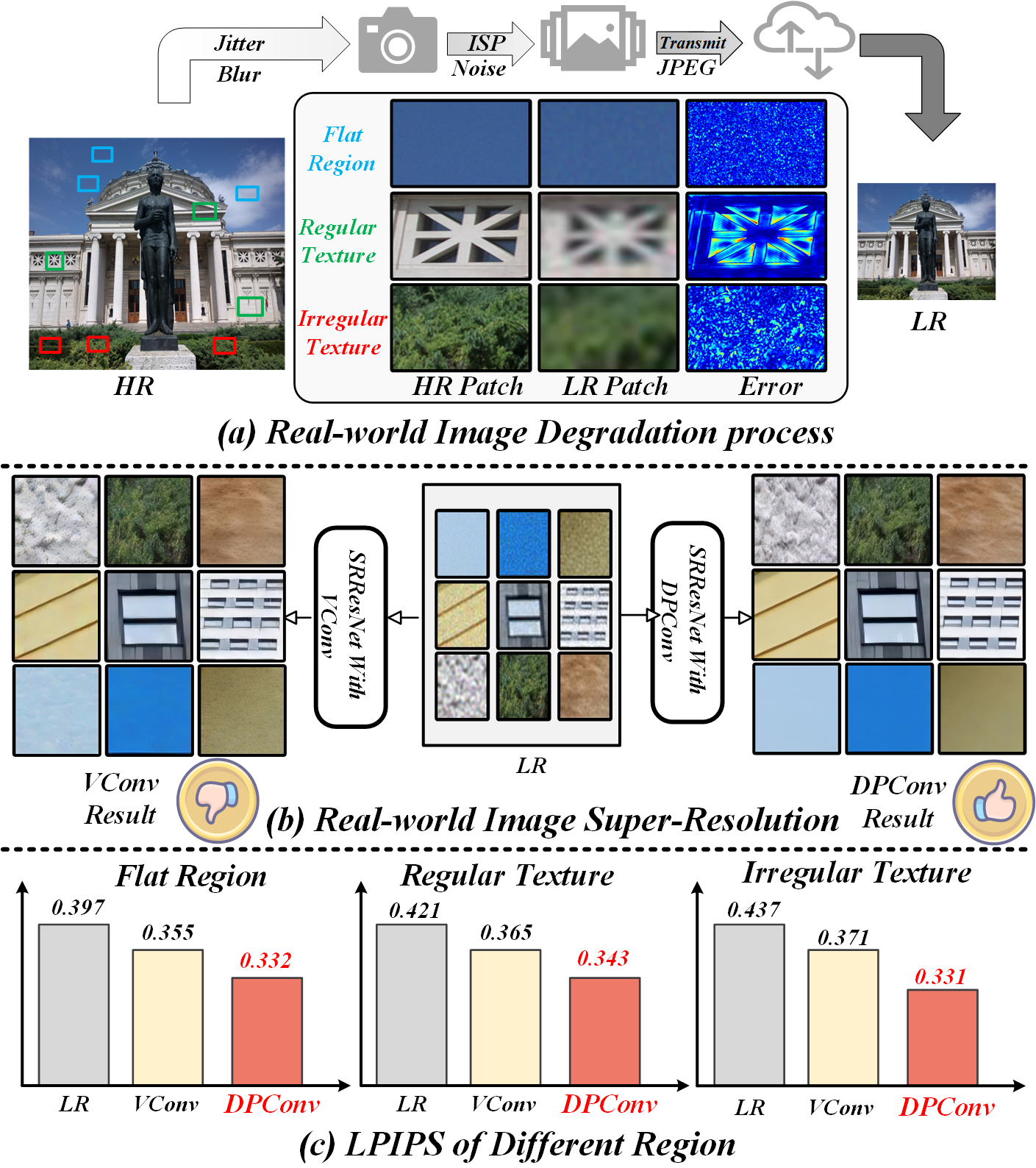}
    \caption{(a) During imaging, real low-resolution (LR) images often undergo various highly nonlinear degradations, leading to inconsistent detail and contrast loss across different regions of the image. Error represents the residuals of HR and LR. (b-c) We introduce DGConv to enhance detail reconstruction and contrast restoration capabilities across different textural and flat regions while maintaining the same inference computational cost as VConv.}
    \label{fig:motivation}
\end{figure}

Real-world image super-resolution (Real-SR) aims to reconstruct high-quality and high-resolution images from low-resolution images captured in real-world scenes \cite{wu2023seesr,niu2024learning,guan2024frequency,wang2023exploiting,zhang2024real}. It has a wide range of applications, from enhancing the visual effect of cameras and smartphones to bolstering the robustness of computer vision systems \cite{shermeyer2019effects,dai2016image}. However, real low-resolution (LR) images often undergo various highly nonlinear degradations. The introduced artifacts disrupt the spatial arrangement characteristics of inherent textures and statistical properties of images \cite{realworld_SR_survey,survey2}, making it difficult to restore image details and contrast, as shown in Fig. \ref{fig:motivation} (a-b). Therefore, for the Real-SR task, the key challenge lies in accurately perceiving the spatial arrangement characteristics of textures across regions of LR images while removing artifacts introduced by degradation clues, achieving high-quality detail restoration \cite{jeelani2023expanding,chen2024iterative,park2023content}.

To solve above challenges, existing work has proposed various well-designed network architectures to enhance image details, which, however, incur high computational costs \cite{li2023efficient,Swinir,sun2023hybrid,liu2023coarse,xia2022knowledge,liang2022efficient,behjati2022frequency,xie2021learning,choi2023n,wei2023taylor}. For example, the real-world image super-resolution method BSRGAN \cite{BSRGAN} and Real-ESRGAN \cite{wang2021real} all adopt cumbersome RRDB as the backbone to reconstruct image details in both regular and irregular texture. Concurrently, to reduce computational costs, many methods have been proposed from the perspective of the frequency domain to improve the capability of modeling high frequencies. For instance, Li \textit{et al.} propose separate images into low, middle, and high frequencies to comprehensively compensate for the information lost for each frequency component \cite{ORNet}. In addition, self-attention has been recently adopted in HAT \cite{HAT} and SwinIR \cite{Swinir} to model the relationship between the different image patches to enhance image details, resulting in a large number of network parameters and FLOPs. Furthermore, to remove the introduced artifacts, some works propose blur estimation networks to estimate the blur kernel of real LR images and utilize estimated blur to guide SR networks to reconstruct image details \cite{IKC,DAN, chen2023better}. From another perspective, researchers propose to utilize Generative Adversarial Networks (GAN) \cite{wang2018esrgan} and perceptual loss \cite{blau20182018} to improve the network’s high-frequency perception ability to obtain a visually pleasing effect that is sensitive to the human visual system. Although existing SR methods achieve impressive achievements in detail recovery, they still fall short when addressing regions with complex gradient arrangements (\textit{e.g.}, irregular texture regions) due to the intensity-based linear weighting feature extraction method in vanilla convolution. Moreover, the stochastic artifacts introduced by degradation cues during the imaging process in real LR can further increase the disorder of the overall image details and statistical properties, which will complicate the intrinsic gradient perception and degrade the image contrast.

To alleviate the above problem, we innovatively introduce kernel-wise differential operations within the convolutional kernel to propose several learnable directional gradient convolutions to provide convolution kernels with directional gradient awareness, including: \textrm{i)} Irregular Directional Gradient convolution (IDG) and {\textrm{ii)}} regular directional gradient convolution, including: Center Surrounding Gradient convolution (CSG), Vertical Gradient convolution (VG), and Horizontal Gradient convolution (HG). In addition, we innovatively introduce aggregation operations within the convolutional kernel to propose learnable Center-Surrounding Aggregation convolution (CSA) to facilitate contrast perception. These convolutions, as well as VConv, are integrated into parallel with a kernel-aware balance mechanism to form an Adaptive Directional Gradient Convolution (DGConv), which balances the enhancement of both regular and irregular textures as well as image contrast. In collaboration with DGConv, we devised an innovative parameter fusion approach for DGConv, preserving its extensive representational abilities while ensuring computational expenses remain on par with a single VConv, which enables DGConv to serve as a ‘plug-and-play’ unit to subplace the VConv in existing SR networks to boost the performance without introducing extra computational costs, as shown in Fig. \ref{fig:motivation} (b-c). Furthermore, to adeptly balance the enhancement of texture and contrast while meticulously investigating the interdependencies between them, we propose an Adaptive Information Interaction Block (AIIBlock) for adaptive selection and fusion of gradient and contrast representations.  Based on DGConv and AIIBlock, we build a simple yet efficient Directional Gradient Perceiving Network (DGPNet) for adaptively enhancing real LR images that achieve state-of-the-art performance with low computational complexity.

The contribution can be summarized as follows:

(1) A novel DGConv is proposed to facilitate the extraction and balance of detail-relevant, contrast-relevant, and degradation-relevant properties by embedding corresponding directional gradient operators into the convolutional process. Moreover, we also propose an equivalent replacement method from VConv to DGConv, which can further improve the performance of the existing SR networks without increasing computational costs.

(2) To adeptly balance the enhancement of texture and contrast while meticulously investigating the interdependencies between them, we propose the Adaptive Information Interaction Block (AIIBlock) in the feature space.

(3) Building upon the DGConv and AIIBlock, we propose a new network architecture, the Directional Gradient Perceiving Network (DGPNet), tailored to effectively enhance real low-resolution (LR) images. Extensive experiments have demonstrated that DGPNet surpasses fifteen existing methods, achieving state-of-the-art performance with commendable efficiency and low computational complexity.

\section{Related Work}
Real-world image super-resolution (Real-SR) aims to reconstruct image details and contrast from the low-resolution versions captured in real scenarios while suppressing the interruption of various degradations. In this paper, we review previous Real-SR works from the following aspects. A more detailed introduction to Real-SR can be found in \cite{wang2020deep,liu2022blind,chen2022real}.

\textbf{Data Generation.} To obtain datasets for training SR networks, researchers propose to collect HR-LR image pairs by adjusting the camera's focal length \cite{chen2019camera,wei2020component,cai2019toward}. However, this collection is labor-intensive, may introduce image misalignment problems, and only covers a few real-world scenarios. To alleviate this issue, some preliminary works propose utilizing the bicubic interpolation method to generate the synthetic training data \cite{SRResNet,EDSR,VDSR,gao2012image,dong2015image,liu2016robust,yang2010image}. However, linear bicubic interpolation is simple and far from the real nonlinear degradation process. Therefore, recent works have considered multiple distortions from camera/object jitter, post-processing by image signal processing (ISP), and high ratio compression during transmission in real scenes and propose a practical and realistic degradation model to simulate the degradation process of real HR-LR image pairs to train the network \cite{BSRGAN,wang2021real,liang2021swinir,xia2023meta,jiang2023fabnet}. As a representative, Wang \textit{et al.} propose a high-order random degradation modeling process with JPEG, blur, noise, downsampling, \textit{etc} to simulate the real degradation process \cite{wang2021real}. In this paper, we follow this degradation model to train and evaluate real-world SR methods. 

In the process of image acquisition, real low-resolution (LR) images suffer from a range of severe nonlinear degradation cues, which results in the loss of fine details and a degradation in contrast quality \cite{realworld_SR_survey,survey2}. Traditional super-resolution techniques, such as bicubic interpolation and SRCNN \cite{dong2016accelerating}, struggle to recover the lost details. To overcome the limitations of existing methods in capturing high-frequency details, recent researches has concentrated on two main strategies: Optimizing Loss Functions \cite{wang2018esrgan,blau20182018}  and Designing Network and Module \cite{xu2022dual,liang2022details,park2023content,yin2023metaf2n,yang2010image,xiao2024ttst,10453456,xu2024uncovering,peng2021ensemble,wang2023decoupling,peng2020cumulative,li2023ntire,yan2023textual,wang2023brightness,peng2024lightweight,he2024latent,ren2024ninth,conde2024real,dai2021feedback,yi2021structure,yi2021efficient}.

\textbf{Optimizing Loss Functions.} Researchers have proposed leveraging Generative Adversarial Networks (GANs) \cite{wang2018esrgan}, perceptual loss functions \cite{blau20182018}, and focal frequency loss \cite{jiang2021focal} to train super-resolution (SR) networks. This approach aims to enhance the network's high-frequency perception capabilities, thereby achieving visually pleasing results for the human visual system. To improve visual perceptual quality in real-world as well as reduce artifacts, researchers propose integrating pixel-level loss functions (\textit{e.g.}, L1 and MSE loss) with the aforementioned losses \cite{liang2022details,wang2021real,BSRGAN,DASR}. For example, Jiang \textit{et al.} propose a loss function in the frequency domain, termed the focal frequency loss, which directly optimizes image reconstruction and synthesis methods within the frequency domain. This approach is designed to enhance the reconstruction of higher-frequency details in the results \cite{jiang2021focal}.

\textbf{Designing Network and Module.} To enhance the network's ability to perceive diverse and complex textures, numerous studies have introduced complex and sophisticated network architectures aimed at improving image details. However, these advanced models often come at a considerable computational cost \cite{li2023efficient,Swinir,sun2023hybrid,liu2023coarse,xia2022knowledge,liang2022efficient,behjati2022frequency,xie2021learning,choi2023n,wei2023taylor}. For instance, state-of-the-art Real-SR methods, such as BSRGAN \cite{BSRGAN} and Real-ESRGAN \cite{wang2021real}, leverage the computationally demanding RRDB (Residual-in-Residual Dense Block) architecture as their foundational backbone for reconstructing image details across both regular and irregular textures. In parallel efforts to mitigate these computational expenses, various approaches have been proposed that focus on the frequency domain to bolster the network's capacity for modeling high-frequency components. For example, Li \textit{et al.} \cite{ORNet} have suggested a method that decomposes images into low, mid, and high-frequency bands to specifically enhance the network's performance in restoring mid- and high-frequency details. Additionally, self-attention mechanisms, which have recently been integrated into models like HAT \cite{HAT} and SwinIR \cite{Swinir}, are utilized to model the interrelationships between different image patches. This strategy, while effective for enhancing detail, also leads to an increase in the number of network parameters and floating-point operations (FLOPs). Moreover, to address the challenges posed by real-world image degradation in real LR, some research has introduced blur estimation networks. These networks estimate the blur kernel of real low-resolution images and use this estimation to guide the super-resolution networks in reconstructing detailed imagery \cite{IKC,DAN, chen2023better}.

Despite the significant advancements in enhancing high-frequency details and contrast through the incorporation of various modules and networks, existing methods face several challenges: i) Current approaches widely utilize VConv to perceive spatial structural information. However, VConv's representation learning, which relies on weighted summation, often falls short in reconstructing texture regions with complex gradient arrangement \cite{wang2020high}. ii) To counteract the effects of degradation cues, many existing methods incorporate dedicated sub-networks \cite{IKC,DAN,chen2023better}. While these additions aim to improve performance, they 
inevitably introduce extra computational overhead. To tackle these issues, we introduce a novel Directional Gradient Convolution (DGConv). Designed as a versatile and easily integrable `plug-and-play' component, DGConv significantly elevates the capabilities of traditional VConv. It effectively models high-frequency image details while also mitigating the disruptive effects of various degradations.

\section{Proposed Method}
In real-world scenarios, real LR images often suffer from various highly nonlinear degradations, leading to inconsistent details and contrast loss across regions of images (flat, regular texture, and irregular texture regions), as shown in Fig. \ref{fig:motivation} (a). Therefore, Real-SR endeavors to produce high-resolution images with rich details while mitigating the impact of various degradation factors. Although existing SR methods based on VConv perform well in mapping from low- to high-resolution images, they often fall short when addressing regions with irregular directional gradient arrangements due to the intensity-based linear weighting feature extraction method. Moreover, the stochastic artifacts introduced during the imaging process in Real LR can further increase the disorder of the overall image gradient and disturb the image's statistical properties, further complicating the extraction of intrinsic gradients and degrading the image contrast. To address this issue, this paper proposes the adaptive Directional Gradient Convolution (DGConv) to enhance the detail reconstruction and contrast representation capabilities of existing VConv in different texture regions and flat regions.

\subsection{The VConv}
Firstly, we review the widely-used VConv $F_{vconv}(X)$ with a kernel size of $3 \times 3$ as follows:

\begin{equation}
F_{vconv}(X)  = \sum\limits_{i = 1}^9 {{\omega_{_i}}}  \cdot {x_i}.
\label{VConv_rewitte}
\end{equation}
where $X$ and $F_{vconv}(X)$ represent input and output feature. We label the $3 \times 3$ square region $L$ from 1 to 9 in a left-to-right and top-to-bottom direction. The feature intensity and learned weights at the location of $L_i$ are represented by $x_{i}$ and $\omega_{i}$, respectively. While existing VConv-based SR methods perform well in mapping from low- to high-resolution images, they often fall short when addressing regions with irregular directional gradient arrangements due to the intensity-based linear weighting feature extraction method. Specifically, it is difficult for VConv to utilize intensity-based linear weighting to extract detail-relevant properties that are primarily contained within the differences between adjacent pixels \cite{ECB,wang2020high}. In addition, degradation cues will further disrupt the spatial arrangement characteristics of inherent textures and statistical properties of images, which makes it hard to enhance image details and contrast.

\subsection{The DGConv}
To alleviate the above problem, we innovatively introduce kernel-wise differential operations within the convolutional kernel to propose several learnable directional gradient units to enhance directional perception, including: {\textrm{i)}} Irregular Directional Gradient convolution (IDG) and {\textrm{ii)}} regular directional gradient convolution, including: Center Surrounding Gradient convolution (CSG), Vertical Gradient convolution (VG), and Horizontal Gradient convolution (HG). In addition, we innovatively introduce aggregation operations within the convolutional kernel to propose learnable Center-Surrounding Aggregation convolution (CSA) to enhance contrast perception. These units, as well as VConv, are integrated in parallel with a kernel-aware balance mechanism to form an Adaptive Directional Gradient Convolution (DGConv), which balances the enhancement of both regular and irregular textures. 

\begin{figure*}[ht]
    \centering
    \includegraphics[width=1.0\linewidth]{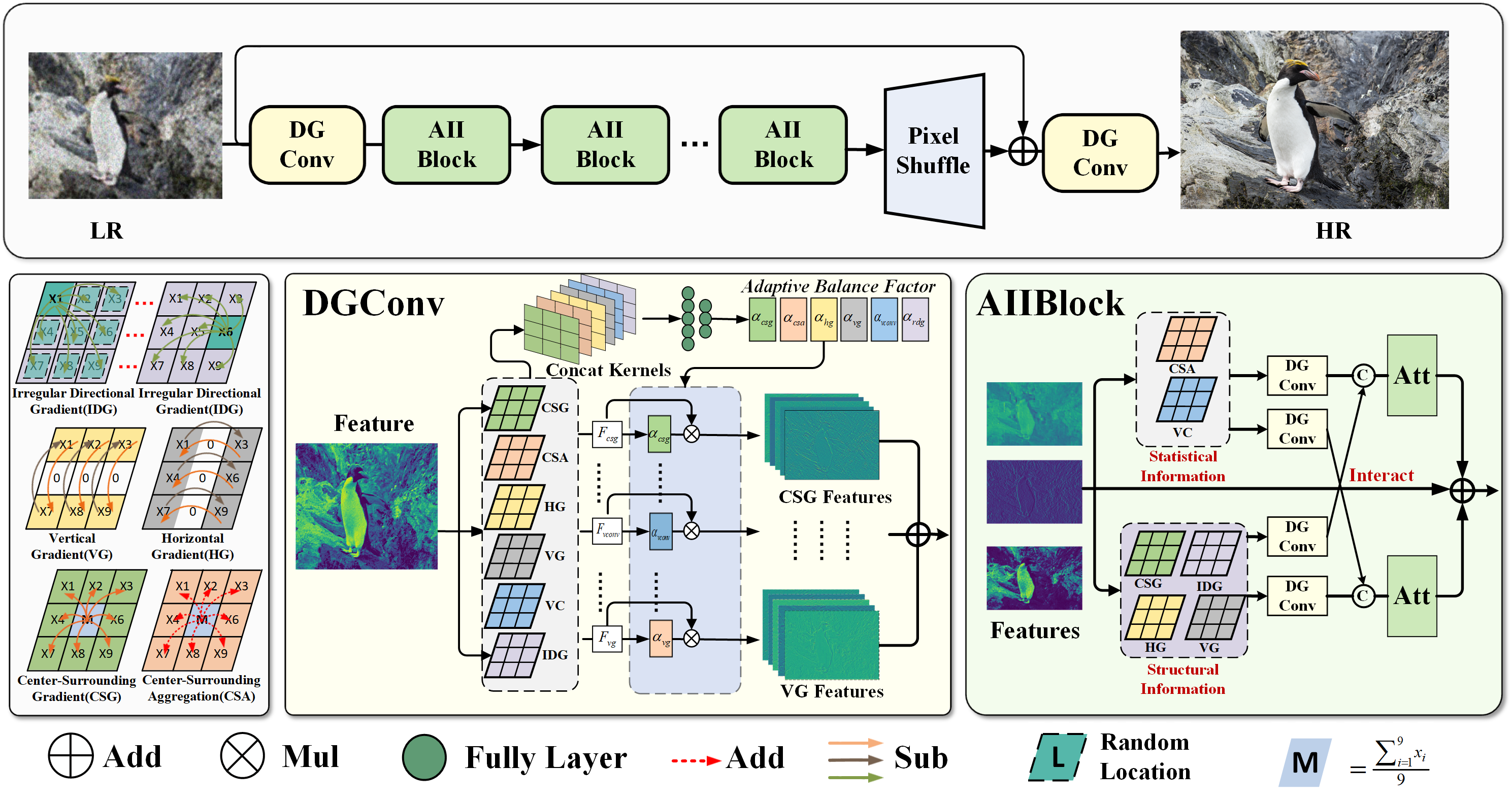}
    \caption{An overview of DGConv and DGPNet. DGConv consists of {\textrm{i})} learnable Irregular Directional Gradient convolution (IDG), {\textrm{ii)} }learnable Regular Directional Gradient convolutions: Center-Surrounding Gradient (CSG) convolution, Vertical Gradient convolution (VG), Horizontal Gradient (HG), \textrm{iii)} Center-Surrounding Aggregation (CSA) convolution and Vanilla Convolution (VConv). Simple yet efficient DGPNet directly stacks $N_{block}$ AIIBlocks as the backbone, uses pixel shuffle to improve the resolution of features, and uses DGConv as image-to-feature and feature-to-image layers.}
    
    \label{fig:Framework}
\end{figure*}

For each IDG, we randomly select a position as the center pixel and then calculate the gradients between the remaining N-1 points and the center point, as shown in Fig. \ref{fig:Framework}. The formulation of the IDG is as follows: 
\begin{equation}
F_{idg}(X) = \sum\limits_{i = 1}^9 {{\omega_{_i}}}  \cdot \left( {{x_i} - {x_j}} \right) \ \ {x_j} \in \{x_1,x_2,...x_n\}.
\label{idg}
\end{equation}
where the definition of $x_{i}$ is the feature intensity of the location of $L_i$, and ${x_j}$ is randomly sampled from all points in the kernel-covered local area. For a layer containing $N$ IDG convolutions with kernel size $K$, $N*K*(K-1)$ directional gradients can theoretically be extracted, which can greatly improve the network's perception of irregular details.

Considering that the human visual system is often more sensitive to regular textures with consistent direction and repeating arrangement (\textit{e.g.}, horizontal, vertical, and diagonal), we propose to utilize difference operation between the local statistical mean and surrounding points to obtain horizontal, vertical, and forty-five-degree gradients to facilitate regular texture perception, termed the Center-Surrounding Gradient (CSG) convolution, as expressed in the following formula:

\begin{equation}
F_{csg}(X) = \sum\limits_{i = 1}^9 {{\omega_{_i}}}  \cdot \left( {{x_i} - \frac{1}{9}  \sum\limits_{i = 1}^9 {{x_{_i}}} } \right).
\label{Laplacian_CNN}
\end{equation}

However, CSG can only extract the gradients of adjacent two pixels. To further improve the richness of gradients, we further perform differential operations between non-adjacent points and propose learnable Horizontal Gradient (HG) convolution and Vertical Gradient (VG) convolution, as expressed in the following:

\begin{equation}
\small
\begin{array}{l}
F_{hg}(X)   = {\omega_{_1}} \cdot \left( {{x_1} - {x_3}} \right) + {\omega_2} \cdot 0 + {\omega_{_3}} \cdot \left( {{x_3} - {x_1}} \right)\\
  \; \; \; \; \; \; \; \; \; \; \; \; \; \; \;+ {\omega_4} \cdot \left( {{x_4} - {x_6}} \right) + {\omega_5} \cdot 0 + {\omega_6} \cdot \left( {{x_6} - {x_4}} \right)\\
\; \; \; \; \; \; \; \; \; \; \; \; \; \; \;+ {\omega_7} \cdot \left( {{x_7} - {x_9}} \right) + {\omega_8} \cdot 0 + {\omega_{_9}} \cdot \left( {{x_9} - {x_7}} \right)
  \\
  F_{vg}(X)  = {\omega_{_1}} \cdot \left( {{x_1} - {x_7}} \right) + {\omega_2} \cdot \left( {{x_2} - {x_8}} \right)+ {\omega_{_3}} \cdot \\ \; \; \; \; \; \;\; \; \; \; \; \;\; \; \; \; \; \; \left( {{x_3} - {x_9}} \right) + {\omega_4} \cdot 0 + {\omega_5} \cdot 0 + {\omega_6} \cdot 0 +  {\omega_7}  \cdot\\ \;\; \; \; \; \; \;\; \; \; \; \; \;   \left( {{x_7} - {x_1}} \right) + {\omega_8} \cdot \left( {{x_8} - {x_2}} \right) + {\omega_{_9}} \cdot \left( {{x_9} - {x_3}} \right).
\end{array}
\label{SobelX}
\end{equation}

Leveraging the proposed IDG, CSG, HG, and VG, we can significantly improve the model's perceptual capabilities in directional gradient extraction. This advancement enables the network to process both regular and irregular texture regions more effectively. Furthermore, these operators can also assist the network in perceiving artifacts introduced by degradation cues, such as horizontal and vertical block artifacts for JPEG compression \cite{pennebaker1992jpeg} and motion artifacts for blurring \cite{pennebaker1992jpeg}. These improvements contribute to the generation of high-quality, high-resolution (HR) images.

Last but not least, considering that degradation cues (\textit{e.g.}, low light \cite{li2021low} and defocus blur \cite{zhang2022deep}) in the real world will reduce image contrast, we innovatively propose the incorporation of an aggregation operation during the convolution process to facilitate contrast-relevant properties modeling. Specifically, we introduce the add operator to suppress the influence of tiny structures to facilitate modeling contrast-relevant properties that mainly utilize pixel intensities and propose a Center-Surrounding Aggregation (CSA) convolution to strengthen the proportion of statistical information needed to recover the statistical properties of images, as expressed in the following formula:
\begin{equation}
F_{csa}(X) = \sum\limits_{i = 1}^9 {{\omega_{_i}}}  \cdot \left( {{x_i} + \frac{1}{9} \sum\limits_{i = 1}^9 {{x_{_i}}} } \right).
\label{AntiLaplacian_CNN}
\end{equation}

In addition, we propose an adaptive fusion mechanism for DGConv to balance the enhancement of both regular and irregular textures while not disrupting the linearity of DGConv to maintain the lightweight of DGConv, as shown in Fig. \ref{fig:Framework}. Specifically, we label the above learnable kernels of irregular gradient, center-surrounding gradient, center-surrounding aggregation, horizontal gradient, vertical gradient, and vanilla convolution as $\omega_{idg}$, $\omega_{csg}$, $\omega_{csa}$, $\omega_{hg}$, $\omega_{vg}$, and $\omega_{vconv}$, respectively. Then, we concatenate them:
\begin{equation}
\omega_{all} = cat(\omega_{idg}, \omega_{csg}, \omega_{csa}, \omega_{hg}, \omega_{vg}, \omega_{vconv}).
\end{equation}

We use the learnable linear layer to learn the adjustment factor $\alpha$ of each convolution branch, as follows:

\begin{equation}
\begin{array}{l}
\alpha_{i}~=~linear_{i}(\omega_{all}) \\
\text{where }{i} \in \{idg,csg,csa,hg,vg,vconv\}.
\end{array}
\end{equation} 

Finally, we add all convolution branches together and form DGConv ${F_{DG}}$, as follows:

\begin{equation}
\begin{array}{l}
{F_{DGConv}} = {\alpha_{idg}} \cdot {F_{idg}} + {\alpha_{csg}} \cdot {F_{csg}}   + {\alpha_{csa}} \cdot {F_{csa}} \\ \; \; \; \;\; \;\; \;\; \; \;\; \;
 + {\alpha_{hg}} \cdot {F_{hg}} + {\alpha_{vg}} \cdot {F_{vg}} + {\alpha_{vconv}} \cdot {F_{vconv}}.\\
\end{array}
\label{DGConv}
\end{equation}

Lastly, the aforementioned adjustment factors are solely dependent on the network's own weights. Before inferencing, these factors can be merged into the network's weights, which reduces the number of parameters and also maintains performance. More details about it can be seen in section \ref{Proposed Equivalent Parameters Fusion}.

\subsection{The DGPNet}

To strike a balance between texture and contrast enhancement while exploring the relationship between them, we propose an Adaptive Information Interaction Block (AIIBlock) for adaptive selection and fusion of gradient and contrast representations, as shown in Fig. \ref{fig:Framework}. Since learnable IDG, HG, VG, and CSG can perceive random and diverse directional gradient information by leveraging the directional arrangement, while VConv and CSA can extract statistical properties by leveraging the statistical and feature response information, we divide gradient-oriented and contrast-oriented information into two branches. Then, we align them, integrate them, and adaptively interact with the information of the different branches. 

Specifically, for the input feature $X$, gradient-oriented $X_{g}$ information is obtained through the IDG, VG, HG, and CSG, and contrast-oriented $X_{c}$ is obtained through the VConv and CSA:
\begin{equation}
\begin{array}{l}
{X_{g}} = cat\left( { {F_{idg}}(X), {F_{csg}}(X),{F_{hg}}(X),{F_{vg}}(X)} \right)\\
{X_{c}} = cat\left( {{F_{vconv}}(X),{F_{csa}}(X)} \right).
\end{array}
\end{equation}
Then, we utilize DGConv to obtain $X_{c1}$ and $X_{g1}$ to align the feature size of another branch and obtain $X_{c2}$ and $X_{g2}$ for the following integration. We concatenate $X_{g1},X_{c2}$ and $X_{c1},X_{g2}$ for information integration between different branches. Finally, we use channel attention SE \cite{SENet} to adaptively select and weight information of each branch and use input $X$ as residual learning to get output $O$, as follows:
\begin{equation}
\small
O = X + SE\left( {cat\left( {{X_{g1}},{X_{c2}}} \right)} \right) + SE\left( {cat\left( {{X_{c1}},{X_{g2}}} \right)} \right).
\end{equation}

Based on those, we build a simple yet efficient Directional Gradient Perceiving Network (DGPNet). DGPNet directly stacks $N_{block}$ AIIBlocks as the backbone without fancy design, uses pixel shuffle to improve feature resolution, and uses DGConv as image-to-feature and feature-to-image layers, as shown in Fig. \ref{fig:Framework}.

\begin{figure}[t]
    \centering
    \includegraphics[width=1.0\linewidth]{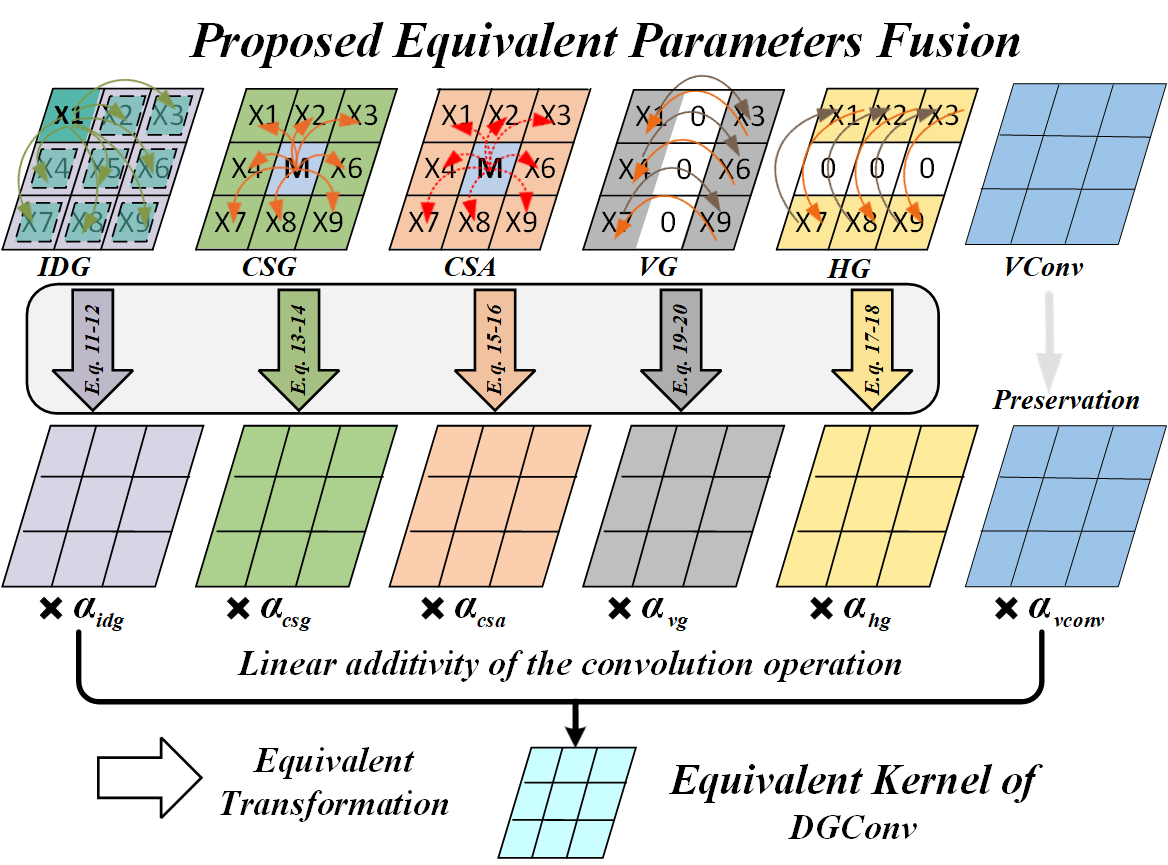}
    \caption{Illustration of our proposed Equivalent Parameters Fusion. It can merge the multiple kernels in DGConv into a single kernel to reduce the computational cost.}
    \label{fig:merge}
\end{figure}

\subsection{Proposed Equivalent Parameters Fusion}
\label{Proposed Equivalent Parameters Fusion}
Despite its excellent representational capabilities, the proposed DGConv inevitably introduces additional computational costs. A straightforward approach is to employ structural re-parameterization techniques \cite{ding2021repvgg,ding2019acnet} to reduce the number of parameters in the network. However, existing structural re-parameterization techniques that are only capable of fusion parameters with multiple VConv, do not apply to our proposed DGConv. To address this issue, we propose a novel equivalent parameter fusion for the DGConv that can simplify its complexity to that of a single VConv during inferencing. Consequently, our equivalent parameter fusion allows DGConv to maintain its rich representational capabilities while keeping computational costs consistent with VConv. Specifically, we first equate the five proposed kernels to the VConv-style and then amalgamate them into a single VConv kernel, as shown in Fig. \ref{fig:merge}. For IDG in Eq. \ref{idg}, we assume that the selected center point ${x_j}$ is ${x_1}$ for the sake of understanding, expand it, combine the coefficients of ${x_i}$, and get a new VConv-style kernel ${\omega}_{idg}^*$, as follows:

\begin{equation}
\begin{array}{l}
{F_{idg}}(X) = \sum\limits_{i = 1}^9 {{\omega_{idg}}}  \cdot ({x_i} - {x_1} )\\ \;\;\;\;\;\;\;\;\;\; \;\;\; = \sum\limits_{i = 1}^9 {\omega_{idg}^*}   \cdot {x_i}.\\
\end{array}
\end{equation}

\begin{equation}
\small
\begin{array}{l}
{\omega^*}_{idg} = \{ {{\omega_{idg}}_{_1} - {\omega_{idg}}_{_1}} ,{\omega_{idg}}_{_2} - {\omega_{idg}}_{_1} , {\omega_{idg}}_{_3} - {\omega_{idg}}_{_1},\\
\; \; \; \; \; \; \; \; \; \; \; \; \; \;\ {{\omega_{idg}}_{_4} - {\omega_{idg}}_{_1}} ,{\omega_{idg}}_{_5} - {\omega_{idg}}_{_1} , {\omega_{idg}}_{_6} - {\omega_{idg}}_{_1},

\\
\; \; \; \; \; \; \; \; \; \; \; \; \; \;\ {{\omega_{idg}}_{_7} - {\omega_{idg}}_{_1}} ,{\omega_{idg}}_{_8} - {\omega_{idg}}_{_1} , {\omega_{idg}}_{_9} - {\omega_{idg}}_{_1} .
\end{array}
\end{equation}

For CSG in Eq. \ref{Laplacian_CNN}, we expand it, combine the coefficients of ${x_i}$, and get a new VConv-style kernel ${\omega}_{csg}^*$, as follows:

\begin{equation}
\begin{array}{l}
{F_{csg}}(X) = \sum\limits_{i = 1}^9 {{\omega_{csg}}}  \cdot ({x_i} - \frac{1}{9}  \sum\limits_{i = 1}^9 {{x_{_i}}} )\\
\;\;\;\;\;\;\;\;\;\;\;\;\; = \sum\limits_{i = 1}^9 {\left( {{\omega_{csg}} - \frac{1}{9}  \sum\limits_{i = 1}^9 {{\omega_{csg}}} } \right)}  \cdot {x_i}\\
\;\;\;\;\;\;\;\;\;\; \;\;\; = \sum\limits_{i = 1}^9 {\omega_{csg}^*}   \cdot {x_i}.\\
\end{array}
\end{equation}

\begin{equation}
{\omega_{csg}^*} = \left( {{\omega_{csg}}} - \frac{1}{9}  \sum\limits_{i = 1}^9 {{\omega_{csg}}} \right).
\end{equation}

In the same way, we can get the equivalent transformation for CSA in Eq. \ref{AntiLaplacian_CNN}, and get a new VConv-style kernel ${\omega}_{csa}^*$, as follows:
\begin{equation}
\begin{array}{l}
{F_{csa}}(X) = \sum\limits_{i = 1}^9 {{\omega_{csa}}}  \cdot ({x_i} - \frac{1}{9}  \sum\limits_{i = 1}^9 {{x_{_i}}} )\\
\;\;\;\;\;\;\;\;\;\;\;\;\; = \sum\limits_{i = 1}^9 {\left( {{\omega_{csa}} - \frac{1}{9}  \sum\limits_{i = 1}^9 {{\omega_{csa}}} } \right)}  \cdot {x_i}\\
\;\;\;\;\;\;\;\;\;\; \;\;\; = \sum\limits_{i = 1}^9 {\omega_{csa}^*}   \cdot {x_i}.\\

\end{array}
\end{equation}

\begin{equation}
{\omega_{csa}^*} = \left( {{\omega_{csa}}} - \frac{1}{9}  \sum\limits_{i = 1}^9 {{\omega_{csa}}}  \right).
\end{equation}

For HG convolution in Eq. \ref{SobelX}, we can translate Eq. \ref{SobelX} and get a new VConv-style kernel ${\omega}_{hg}^*$, as follows:

\begin{equation}
F_{hg}(X) = \sum\limits_{i = 1}^9 {{\omega^*}_{hg}}  \cdot \left( {{x_i}} \right).
\end{equation}

\begin{equation}
\begin{array}{l}

{\omega^*}_{hg} = \{ \left( {{\omega_{hg}}_{_1} - {\omega_{hg}}_{_3}} \right),0,\left( {{\omega_{hg}}_{_3} - {\omega_{hg}}_{_1}} \right),\\
\; \; \; \; \; \; \; \; \; \; \; \; \;\; \; \; \ \left( {{\omega_{hg}}_{_4} - {\omega_{hg}}_{_7}} \right),0,\left( {{\omega_{hg}}_{_7} - {\omega_{hg}}_{_3}} \right)

\\ \; \; \; \; \; \; \; \; \; \; \; \; \;\; \; \; \;
\left( {{\omega_{hg}}_{_7} - {\omega_{hg}}_{_9}} \right),0,\left( {{\omega_{hg}}_{_9} - {\omega_{hg}}_{_7}} \right)\} .
\end{array}
\end{equation}
For VG in Eq. \ref{SobelX}, we can translate Eq. \ref{SobelX} and get a new VConv-style kernel ${\omega}_{vg}^*$, as follows:

\begin{equation}
{F_{vg}}(X) =  \sum\limits_{i = 1}^9 {{\omega^*}_{vg}}  \cdot \left( {{x_i}} \right).
\end{equation}

\begin{equation}
\begin{array}{l}

{\omega^*}_{vg} = \{ ({{\omega_{vg}}_{_1} - {\omega_{vg}}_{_7}} ), ( {{\omega_{vg}}_{_2} - {\omega_{vg}}_{_8}} ),\\ 
\; \; \; \; \; \; \; \; \; \; \; \; \; \; \; \;
( {{\omega_{vg}}_{_3} - {\omega_{vg}}_{_9}} ),
\; \; \; \; \; \; \; 0,
\; \; \; \; \; \; \; 0 \\
\; \; \; \; \; \; \; \; \; \; \; \; \; \; \; \; 0,
\; \; \; \; \; \; \; \; \; \; \; \; \; \; \; \;( {{\omega_{vg}}_{_7} - {\omega_{vg}}_{_1}} ),\\ 
\; \; \; \; \; \; \; \; \; \; \; \; \; \; \; \;
 ( {{\omega_{vg}}_{_8} - {\omega_{vg}}_{_2}} ),( {{\omega_{vg}}_{_9} - {\omega_{vg}}_{_3}} )\}. 
\end{array}
\end{equation}

Then, we can obtain the equivalent VConv-style kernel ${\omega}_{idg}^*$, ${\omega}_{csg}^*$, ${\omega}_{csa}^*$, ${\omega}_{vg}^*$, ${\omega}_{hg}^*$ of learnable Random Direction Gradient, Center-Surrounding Gradient, Center-Surrounding Aggregation, Vertical Gradient, Horizontal Gradient convolutions, respectively. With the linear additivity of convolution, we fuse all kernels into a single final VConv-kernel ${\omega}_{f}^*$, as follows:

\begin{equation}
\begin{array}{l}
\begin{array}{*{20}{l}}
{F_{dg}}(X) = {\alpha_{csg}} \cdot {F_{csg}}(X) + {\alpha_{VConv}} \cdot {F_{vconv}}(X) +  \\ \;\;\;\;\;\;  \;\;\; \;\;\;\;\;\;\;
{\alpha_{csa}} \cdot   {F_{csa}}(X) + {\alpha_{hg}} \cdot {F_{hg}}(X) + 
\\ \;\;\;\;\;\;  \;\;\; \;\;\;\;\;\;\; {\alpha_{rgd}} \cdot {F_{rgd}}(X) + {\alpha_{vg}} \cdot {F_{vg}}(X)
\\
\;\;\;\;\;\;  \;\;\; \;\;\;
= \sum\limits_{i = 1}^9 {\omega_f^*}  \cdot {x_i}.
\end{array}
\end{array}
\end{equation}

\begin{equation}
\begin{array}{l}
{\omega^*}_f = {\alpha_{VConv}}\cdot {\omega_{VConv}} + {\alpha_{lap}} \cdot {\omega^*}_{lap} + {\alpha_{csa}} \cdot {\omega^*}_{csa}\\
\;\;\;\;\;\;\; + {\alpha_{hg}} \cdot {\omega^*}_{hg} + {\alpha_{vg}} \cdot {{\omega^*}_{vg}}+{\alpha_{idg}} \cdot {{\omega^*}_{idg}}.
\end{array}
\end{equation}

Finally, we can equate the kernels of the proposed DGConv to single kernel ${\omega}_{f}^*$ with the same computational complexity as common VConv to reduce the computational cost of DGPNet while effectively extracting various representations in real LR images.

\subsection{Training Loss}
Following previous works \cite{BSRGAN,wang2021real}, we use the L1 loss $L_{l}$, the perceptual loss $L_{p}$, and the adversarial loss $L_{a}$ for training our methods. The total loss is as follows:

\begin{equation}
L_{total} = {\lambda_{l}} \cdot L_{l} + {\lambda_{p}} \cdot L_{p} + {\lambda_{a}} \cdot L_{a}.
\end{equation}
Here ${\lambda_{l}}$, $\lambda_{p}$ and ${\lambda_{a}}$ represent the balancing parameters, which are set to $1.0$, $1.0$ and $0.1$, respectively.

\begin{table*}[t]
\centering
\caption{Comparison with existing methods. Avg represents the average performance of all scenarios. The best and the second-best average performance are marked as {\color[HTML]{FF0000} {\textbf{bold}}} and {\color[HTML]{0000FF} \textbf{bold}}, respectively.}
\resizebox{\linewidth}{!}{%
\begin{tabular}{c||cccccccccc||cc}
\hline
Data     & \multicolumn{2}{c}{Level-I} & \multicolumn{2}{c}{Level-II} & \multicolumn{2}{c}{Level-III} & \multicolumn{2}{c}{Level-IV} & \multicolumn{2}{c||}{Level-V} & \multicolumn{2}{c}{Avg}                                                        \\ 
Metrics  & PSNR$\uparrow$         & LPIPS$\downarrow$       & PSNR$\uparrow$          & LPIPS$\downarrow$        & PSNR$\uparrow$          & LPIPS$\downarrow$         & PSNR$\uparrow$          & LPIPS$\downarrow$        & PSNR$\uparrow$          & LPIPS$\downarrow$        & PSNR$\uparrow$                                  & LPIPS$\downarrow$                                  \\ \hline
SRResNet \cite{SRResNet} & 26.72        & 0.223        & 26.64         & 0.258        & 24.17         & 0.402         & 23.82         & 0.427        & 23.63         & 0.447        & 25.00                                 & 0.351                                 \\
EDSR \cite{EDSR}     & 26.67        & 0.238        & 26.59         & 0.266        & 24.13         & 0.408         & 23.80         & 0.430        & 23.61         & 0.449        & 24.96                                 & 0.358                                 \\
RRDB \cite{wang2018esrgan}   & 26.40        & 0.231        & 26.34         & 0.261        & 23.63         & 0.403         & 23.33         & 0.429        & 23.19         & 0.447        & 24.58                                 & 0.354                                 \\
VDSR \cite{VDSR}   & 24.35        & 0.304        & 24.22         & 0.353        & 23.15         & 0.501         & 22.88         & 0.527        & 22.73         & 0.537        & 23.47                                 & 0.444                                 \\
SiwnIR \cite{Swinir}  & 26.34        & 0.210        & 26.21         & 0.239        & 24.05         & 0.395         & 23.73         & 0.420        & 23.52         & 0.441        & 24.77                                 & 0.341                                 \\
HAT \cite{HAT},     & 27.03        & 0.206        & 26.89         & 0.237        & 24.55         & 0.380         & 24.21         & 0.406        & 23.97         & 0.427        & 25.33                                 & 0.331                                 \\
ESRGAN \cite{wang2018esrgan}  & 21.20        & 0.472        & 22.01         & 0.525        & 23.83         & 0.708         & 23.46         & 0.728        & 23.33         & 0.719        & 22.76                                 & 0.630                                 \\
ESRGAN-FS \cite{ESRGAN-FS}      & 23.64        & 0.359        & 23.09         & 0.457        & 23.48         & 0.632         & 23.20         & 0.655        & 23.04         & 0.671        & 23.29                                 & 0.554                                 \\
GFSNet \cite{GFSNet}  & 25.84        & 0.332        & 25.52         & 0.437        & 24.30         & 0.722         & 23.92         & 0.758        & 23.70         & 0.748        & 24.66                                 & 0.599                                 \\
DAN \cite{DAN}     & 27.25        & 0.336        & 26.78         & 0.460        & 24.31         & 0.748         & 23.90         & 0.768        & 23.69         & 0.750        & 25.19                                 & 0.612                                 \\
IKC \cite{IKC}     & 25.05        & 0.362        & 24.98         & 0.471        & 22.69         & 0.723         & 22.33         & 0.746        & 22.48         & 0.736        & 23.50                                 & 0.607                                 \\
Real-ESRGAN \cite{wang2021real} & 26.13        & 0.231        & 26.08         & 0.240        & 24.10         & 0.343         & 23.79         & 0.368        & 23.44         & 0.382        & 24.71                                 & {\color[HTML]{0000FF} \textbf{0.312}} \\
BSRGAN \cite{BSRGAN}  & 26.76        & 0.240        & 26.64         & 0.250        & 24.25         & 0.372         & 23.91         & 0.404        & 23.77         & 0.405        & 25.07                                 & 0.334                                 \\
DASR \cite{DASR}    & 26.68        & 0.350        & 26.20         & 0.471        & 23.90         & 0.755         & 23.49         & 0.777        & 23.38         & 0.758        & 24.73                                 & 0.622                                 \\
HGGT \cite{HGGT}    & 26.85        & 0.179        & 26.91         & 0.229        & 24.42         & 0.710         & 24.03         & 0.726        & 23.81         & 0.721        & 25.20                                 & 0.513                                 \\ \hline \rowcolor[HTML]{EFEFEF} 
\textbf{DGPNet\_tiny}     & 27.18        & 0.221        & 27.08         & 0.255        & 24.72         & 0.385         & 24.33         & 0.414       & 24.04         & 0.433        & {\color[HTML]{0000FF} {\textbf{25.37}}} & 0.335                                 \\ \rowcolor[HTML]{EFEFEF} 
\textbf{DGPNet}     & 27.35        & 0.201        & 27.26         & 0.231        & 24.76         & 0.362         & 24.38         & 0.385        & 24.15         & 0.405        & {\color[HTML]{FF0000} \textbf{25.41}} & {\color[HTML]{FF0000} {\textbf{0.309}}} \\ \hline
\end{tabular}
}
\label{tab:Compare_with_existing_method}
\end{table*}
\begin{table*}[t]
\centering
\caption{Comparison with existing methods on real-world datasets \cite{cai2019toward}. Our proposed method achieves the best performance.}
\resizebox{\linewidth}{!}{

\begin{tabular}{cc||cccccccccccc
>{\columncolor[HTML]{FFFFFF}}c }
\hline
                        & Metrics & SRResnet  &HGGT & SwinIR & ESRGAN & GFSNet & DAN   & IKC   & RESRGAN & BSRGAN & DASR  & RRDB & \textbf{DGPNet}   \\ \hline
                        & PSNR$\uparrow$    & 28.11    & 27.95 & 26.84  & 27.73  & 27.90  & 27.99 & 27.51 & 26.06   & 26.90  & 27.97 & \textcolor{blue}{\textbf{28.31}} & \textcolor{red}{\textbf{28.57}} \\
\multirow{-2}{*}{Canon} & LPIPS$\downarrow$   & 0.201    & 0.355 & 0.210  & 0.415  & 0.406  & 0.410 & 0.321 & 0.261   & 0.257  & 0.407 & \textcolor{blue}{\textbf{0.195}} & \textcolor{red}{\textbf{0.194}} \\ \hline
                        & PSNR$\uparrow$    & 27.54    & 27.41 & 26.55  & 27.28  & 27.50  & 27.61 & 26.85 & 25.62   & 26.11  & 27.60 & \textcolor{blue}{\textbf{27.86}}  & \textcolor{red}{\textbf{28.06}} \\
\multirow{-2}{*}{Nikon} & LPIPS$\downarrow$   & 0.242    & 0.355 & 0.246  & 0.413  & 0.427  & 0.413 & 0.340 & 0.285   & 0.280  & 0.408 & \textcolor{blue}{\textbf{0.228}} & \textcolor{red}{\textbf{0.222}} \\ \hline
\end{tabular}
}

\label{tab:Real_data}
\end{table*}
\begin{table*}[t]
\centering

\caption{Cost-free improvement for our DGConv during inferencing. VConv and DGConv version models are marked as bold and {bold*}, and the average performance of them are marked as \textbf{bold} and {\color[HTML]{FF0000} \textbf{bold}}, respectively. \ul{We can observe that using DGConv for the same baseline can significantly outperform VConv. It is worth noting that after using our equivalent transformation, the DGConv-based and the VConv-based methods have exactly the same structure and computational complexity but with better performance.} }

\resizebox{\linewidth}{!}{

\begin{tabular}{cc||cc||cc||cc||cc||cc}
\Xhline{2.\arrayrulewidth}
Data                                          & Metrics & VDSR\cite{VDSR}            & {VDSR*}                      & SRResNet\cite{SRResNet}        & {SRResNet*}                    & IMDN\cite{IMDN}            &{IMDN*}                         &EDSR            &{EDSR*}                         & PAN\cite{PAN}             &{PAN*}                          \\ \hline
                                              & PSNR    & 24.58          & 25.40                                 & 26.72          & 26.77                                 & 25.72          & 26.18                                 & 26.73          & 26.83                                 & 26.06          & 26.21                                 \\
\multirow{-2}{*}{Level-I}                     & LPIPS   & 0.298           & 0.274                                  & 0.223           & 0.211                                  & 0.235           & 0.221                                  & 0.234           & 0.230                                  & 0.274           & 0.266                                  \\
                                              & PSNR    & 24.45          & 25.27                                 & 26.64          & 26.74                                 & 25.49          & 26.03                                 & 26.58          & 26.73                                 & 25.98          & 26.09                                 \\
\multirow{-2}{*}{Level-II}                    & LPIPS   & 0.348           & 0.315                                  & 0.258           & 0.243                                  & 0.270           & 0.256                                  & 0.263           & 0.258                                  & 0.305           & 0.297                                  \\
                                              & PSNR    & 23.32          & 23.63                                 & 24.17          & 24.15                                 & 23.36          & 23.76                                 & 24.17          & 24.30                                 & 23.83          & 23.91                                 \\
\multirow{-2}{*}{Level-III}                   & LPIPS   & 0.488           & 0.460                                  & 0.402           & 0.392                                  & 0.411           & 0.411                                  & 0.402           & 0.396                                  & 0.445           & 0.440                                  \\
                                              & PSNR    & 23.03          & 23.30                                 & 23.81          & 23.80                                 & 23.03          & 23.35                                 & 23.84          & 23.98                                 & 23.51          & 23.58                                 \\
\multirow{-2}{*}{Level-IV}                    & LPIPS   & 0.513           & 0.486                                  & 0.427           & 0.414                                  & 0.437           & 0.441                                  & 0.424           & 0.418                                  & 0.468           & 0.465                                  \\
                                              & PSNR    & 22.87          & 23.14                                 & 23.63          & 23.62                                 & 22.93          & 23.24                                 & 23.63          & 23.76                                 & 23.34         & 23.41                                 \\
\multirow{-2}{*}{Level-V}                     & LPIPS   & 0.522           & 0.499                                  & 0.447           & 0.435                                  & 0.456           & 0.460                                  & 0.444           & 0.440                                  & 0.483           & 0.483                                  \\ \hline
\rowcolor[HTML]{EFEFEF} 
\cellcolor[HTML]{EFEFEF}                      & PSNR    & \textbf{23.65} & {\color[HTML]{FF0000} \textbf{24.14}} & \textbf{24.99} & {\color[HTML]{FF0000} \textbf{25.06}} & \textbf{24.11} & {\color[HTML]{FF0000} \textbf{24.52}} & \textbf{24.99} & {\color[HTML]{FF0000} \textbf{25.18}} & \textbf{24.54} & {\color[HTML]{FF0000} \textbf{24.69}} \\
\rowcolor[HTML]{EFEFEF} 
\multirow{-2}{*}{\cellcolor[HTML]{EFEFEF}Avg} & LPIPS   & \textbf{0.434}  & {\color[HTML]{FF0000} \textbf{0.408}}  & \textbf{0.352}  & {\color[HTML]{FF0000} \textbf{0.335}}  & \textbf{0.362}  & {\color[HTML]{FF0000} \textbf{0.354}}  & \textbf{0.353}  & {\color[HTML]{FF0000} \textbf{0.343}}  & \textbf{0.395}  & {\color[HTML]{FF0000} \textbf{0.386}}  \\ \hline
\end{tabular}
}
\label{tab:cost_free_improvement}
\end{table*}

\section{Experiments and Analysis}
\subsection{Experiments Settings}

\textbf{Training Details.} We follow previous works 
\cite{wang2021real,wang2018esrgan} and use the DIV2K \cite{DIV2k}, Flickr2K \cite{Flickr2K}, and OutdoorScene-Training \cite{OutdoorScene-Training} as HR and utilize the degradation model \cite{wang2021real,BSRGAN} to synthetic paired real data to train our proposed method. The networks are trained with the Adam optimizer with a learning rate of $0.0001$. The number of middle feature channels in DGPNet and DGPNet\_tiny is set to 64 and 32, respectively. We set the batch size and training iteration to 16 and 200K and set the scale factor of SR, $N_{block}$, and LR patch size to 4, 16, and 64 × 64 for training on eight NVIDIA RTX3090 GPUs at Pytorch \cite{paszke2019pytorch}.

\textbf{Metrics.} Following previous works \cite{wang2021real,BSRGAN}, we use LPIPS \cite{zhang2018unreasonable} and PSNR \cite{huynh2008scope} to quantitatively evaluate the performance. Note that the higher the PSNR, the better, and the lower the LPIPS, the better. Since the model computation cost is also critical for the Real-SR method, we use FLOPs, Parameters, and running time to evaluate model complexity.

\textbf{Evaluation.}
Following the approach of \cite{wang2021real,BSRGAN}, we synthesize 500 low-resolution and high-resolution image pairs by applying five different levels of degradation for evaluation. And we use the real-world DSLR dataset \cite{cai2019toward}, which contains 459 scenes for training and 100 scenes for evaluation. In addition, we also utilize the real internet dataset RealWorld38 \cite{BSRGAN,wang2021real} for evaluation.

\textbf{Comparisons with State-of-the-art Methods.}
We compare our model with fifteen state-of-the-art Real-SR methods, including SRResNet \cite{SRResNet}, EDSR \cite{EDSR}, RRDB \cite{wang2018esrgan}, VDSR \cite{VDSR}, SwinIR \cite{Swinir}, HAT \cite{HAT}, DAN \cite{DAN}, IKC \cite{IKC}, Real-ESRGAN \cite{wang2021real}, BSRGAN \cite{BSRGAN}, ESRGAN \cite{wang2018esrgan}, ESRGAN-FS \cite{ESRGAN-FS}, GFSNet \cite{GFSNet}, DASR \cite{DASR}, and HGGT \cite{HGGT}. For widely-used image super-resolution (SR) baselines, we retrain them in the same manner as our method for a fair comparison, including SRResNet, EDSR, RRDB, VDSR, SwinIR, DARSR, and HAT. For real-world image super-resolution methods that mainly adopt RRDB as their backbone, we evaluate them using the publicly available best models from their official code for a fair comparison, including ESRGAN, ESRGAN-FS, GFSNet, BSRGAN, DASR, Real-ESRGAN, and HGGT. Additionally, some methods cannot be retrained on our degradation models because they require the blur kernel of real LR images during training, which is unavailable in hybrid degradation scenarios. For these methods, we use the publicly available best models from their official code for evaluation, including IKC and DAN.

\begin{figure}[tp]
    \centering
    \includegraphics[width=1.0\linewidth]{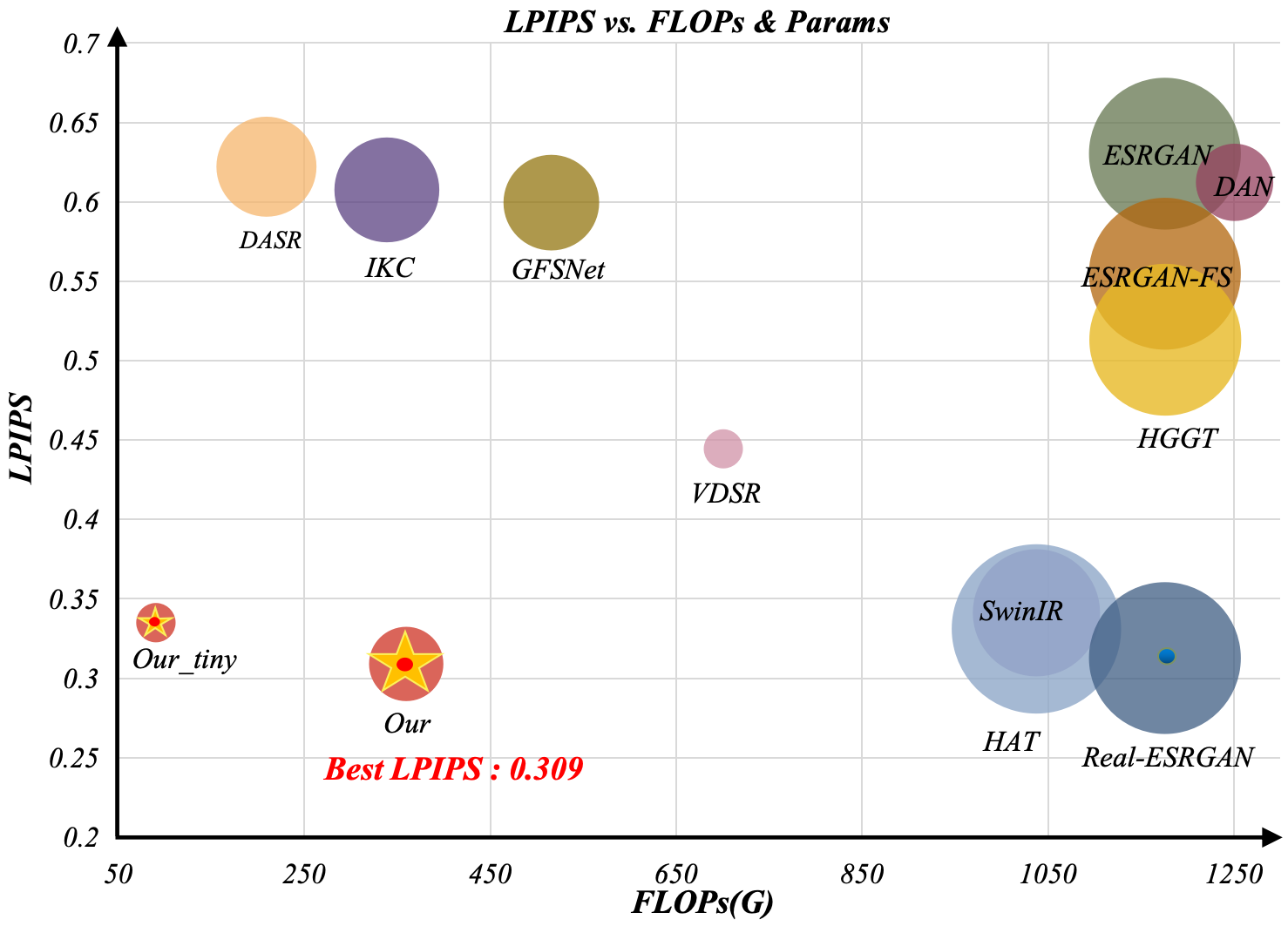}
    \caption{Comparison of model complexity and performance.  Our approach achieves superior performance with fewer parameters and lower FLOPs. Note that a lower LPIPS value indicates better performance. The size of the circle represents the number of parameters.}
    \label{fig:PNSR_Param_FLOPs}
\end{figure} 

\begin{table}[t]
\centering
\caption{Comparison of running time on RTX 3080 GPU. We run 1000 images with $3 \times 128 \times 128$ resolution on $\times 4$ SR and take the average running time as the result.}
\resizebox{\columnwidth}{!}{
\begin{tabular}{c|ccccc}
\Xhline{2.\arrayrulewidth}
         & DAN    & VDSR   & BSRGAN & Real-ESRGAN & HGGT   \\ \hline
Time(s)$\downarrow$ & 0.0665 & 0.0223 & 0.0521 & 0.0522      & 0.0523 \\ \hline

         & SwinIR & ESRGAN & HAT    & RRDB        & \textbf{DGPNet} \\ \hline
Time(s)$\downarrow$  & 0.0475 & 0.0521 & 0.1953 & 0.0523      & \textbf{0.0204} \\ \Xhline{2.\arrayrulewidth}
\end{tabular}}
\label{tab:running_time}
\end{table}

\subsection{Quantitative Results}
\textbf{The performance of DGPNet.} We compare our DGPNet with fifteen Real-SR methods, and the results are shown in Table. \ref{tab:Compare_with_existing_method} and Table. \ref{tab:Real_data}. Our proposed method can achieve the best performance on PSNR and LPIPS and significantly outperform the existing SOTA method (\emph{i.e.}, HAT) by 0.07 dB and 0.022 on PSNR and LPIPS. It proves that our method performs better in real-world scenarios. More importantly, our method has significant advantages over the existing SOTA methods in terms of FLOPs, parameters, and running time, as shown in Table. \ref{tab:running_time} and Fig. \ref{fig:PNSR_Param_FLOPs}. The parameters of our method only account for 21.5\% of HAT and 26.8\% of Real-ESRGAN, and the FLOPS of our method only account for 34.6\% of HAT and 30.5\% of Real-ESRGAN, which demonstrate that our method not only has better performance but also has the advantage of being lightweight and having a low computational cost, making it better suited for real-scenario applications. We further propose a tiny version of our network DGPNet\_tiny with only 32 feature channels, which accounts for only 8.7\% and 7.6\% of the FLOPs of HAT and Real-ESRGAN and has achieved comparable performance. In addition, to evaluate the generalization ability of the proposed method in real-world DSLR camera scenarios, we evaluate our method on the dataset proposed by \cite{cai2019toward}, as shown in Table. \ref{tab:Real_data}. We can observe that our method achieves the best performance, demonstrating our superiority in real-world scenarios.

\textbf{The Generalizability of DGConv.} We use DGConv to replace the VConv in the five classical SR methods, including VDSR \cite{VDSR}, SRResNet \cite{SRResNet}, IMDN \cite{IMDN}, EDSR \cite{EDSR} and PAN \cite{PAN}, and perform training procedure with the same experimental settings. In Table. \ref{tab:cost_free_improvement}, we can observe that the performance of VConv based baseline is comprehensively improved after utilizing the DGConv. For example, with DGConv, the performance of VDSR can be improved by 0.496 dB and 0.026 on PSNR and LPIPS, respectively. This demonstrates that DGConv can be applied to various networks and plugged into existing networks to improve performance.

\subsection{Qualitative Results}
\begin{figure*}[h]
    \centering
    \includegraphics[width=1.0\linewidth]{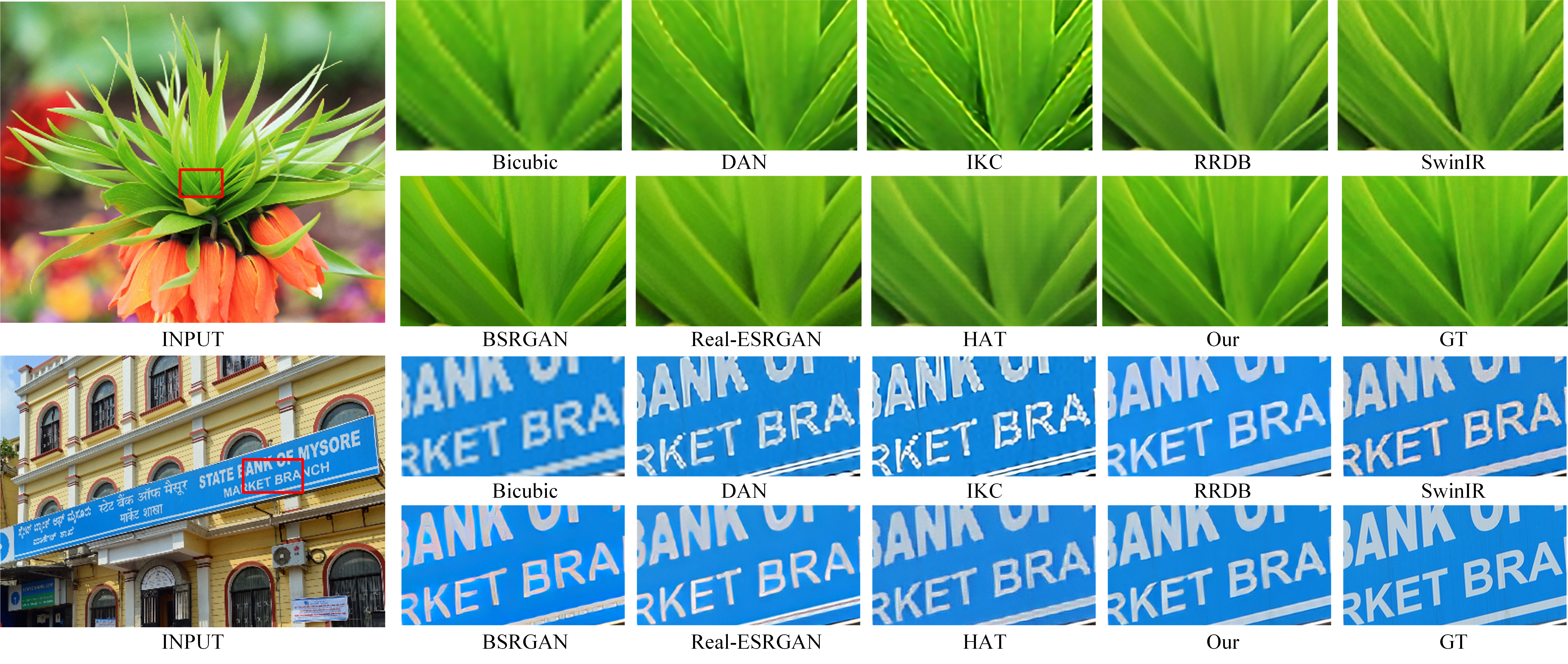}
    \caption{Visual comparison with existing Real-SR methods. Our method achieves better detail recovery and contrast enhancement results. Please zoom in for better visualization.}
    
    \label{fig:compared_image_SR}
\end{figure*}

Fig. \ref{fig:compared_image_SR} shows the qualitative comparison between DGPNet and existing methods. We can observe that, compared with existing methods, our method achieves better image details and contrast in the flat regions and regular and irregular textures. For example, as shown in the regular textures area of the second scene in Fig. \ref{fig:compared_image_SR}, DAN \cite{DAN} and IKC \cite{IKC} introduce extra artifacts and fail to restore the image textures. The outcomes of HAT \cite{HAT}, RRDB \cite{wang2018esrgan}, and SwinIR \cite{Swinir} can both lead to the bending of structural textures, such as those found in the text region. However, our method, benefiting from its perception capacity of directional gradient, preserves a better visual representation of the text's structure.

\begin{figure*}[h]
    \centering
    \includegraphics[width=1.0\linewidth]{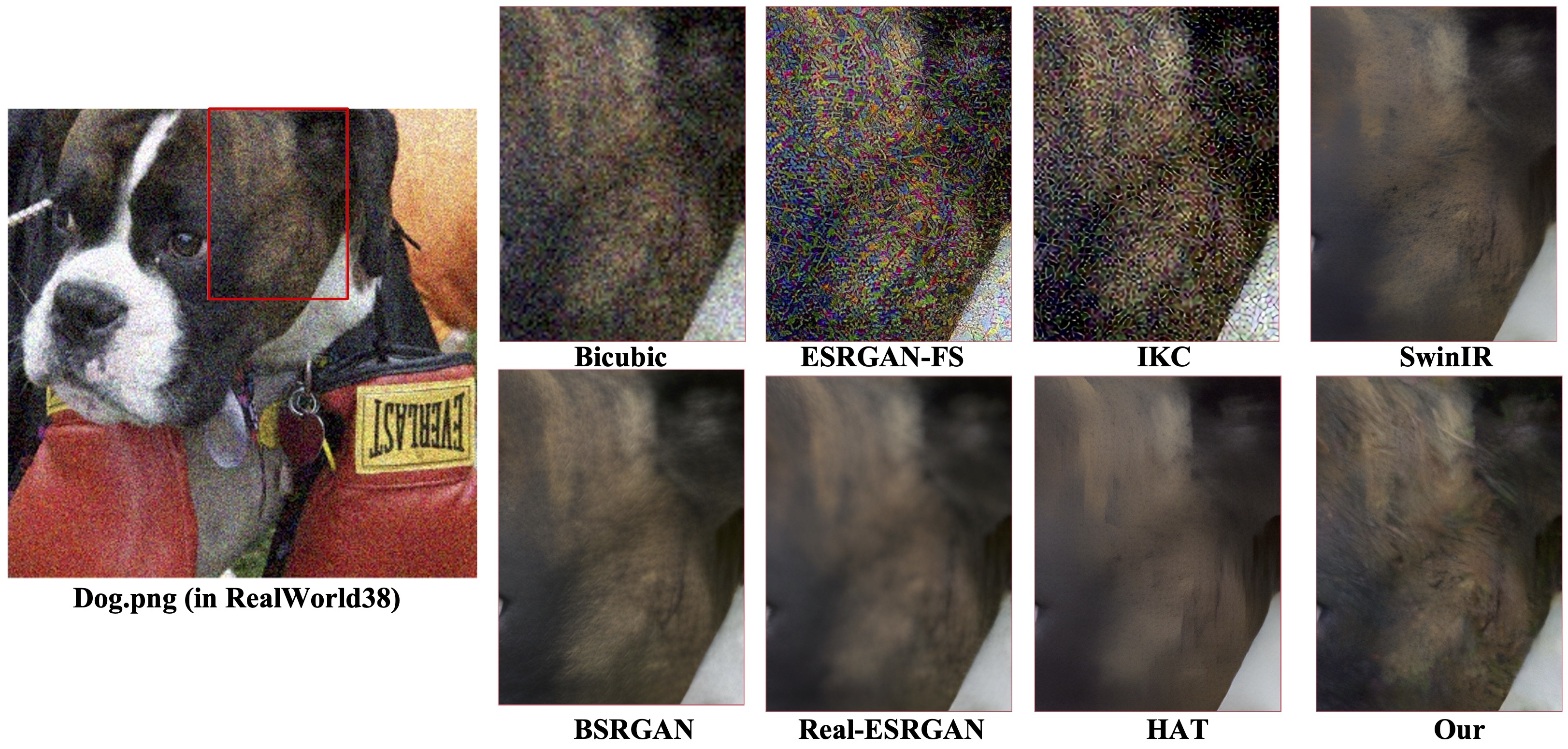}
    \caption{Visual comparisons on real-world data sets RealWorld38 \cite{BSRGAN,wang2021real}. Our method not only removes artifacts more cleanly but also restores better image detail in irregular hair area. Please zoom in for better visualization.}
    \label{fig:real1}
\end{figure*}

In addition, we employ the wild-used real-world data sets RealWorld38 \cite{BSRGAN,wang2021real} for qualitative comparisons, where the images are degraded by real degradation in the complex real world. As shown in Fig. \ref{fig:real1}, we can observe that ESRGAN-FS \cite{ESRGAN-FS} and IKC \cite{IKC} are hard to reconstruct image details, introduce extra artifacts, and further aggravate image damage. And, Real-ESRGAN \cite{wang2021real}, HAT \cite{HAT}, and SwinIR \cite{Swinir} struggle to perceive the irregular direction arrangement of textures in an image, which can lead to an overly smooth appearance, resulting in the loss of image details. However, we can observe that our method better reconstructs image details in irregular textures and better enhances contrast, demonstrating our method's superiority in visual effects in real scenes.

\begin{figure*}[t]
    \centering
    \includegraphics[width=1.0\linewidth]{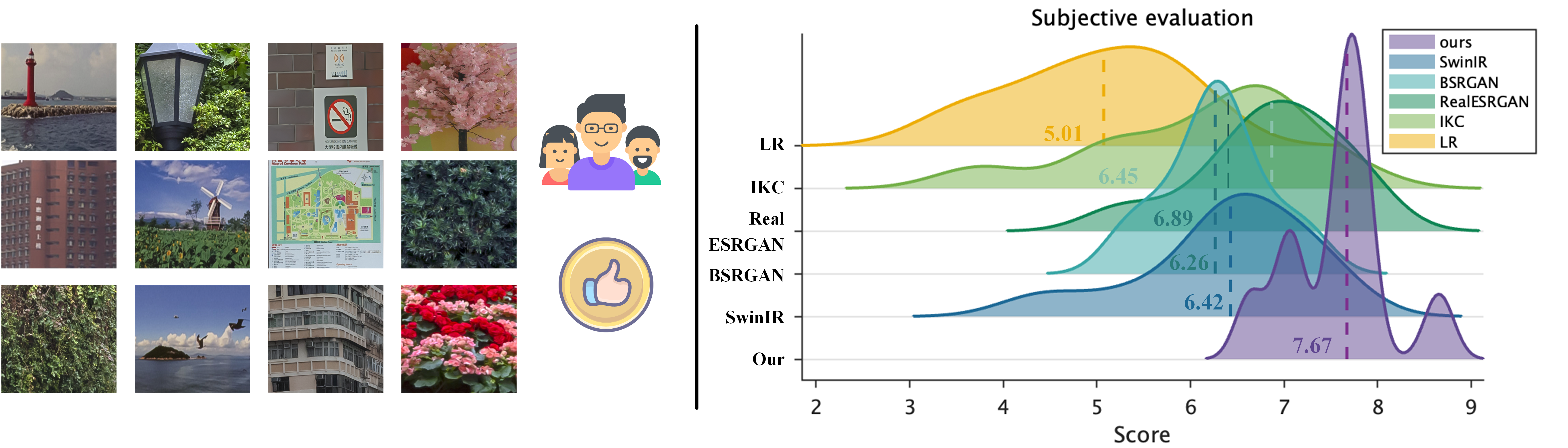}
    \caption{The left is the examples of various real LR images. The right is the user study on the visual perceptual quality of results on real images.}
    \label{fig:User_study}
\end{figure*}
\begin{figure*}[h]
    \centering
    \includegraphics[width=1.0\linewidth]{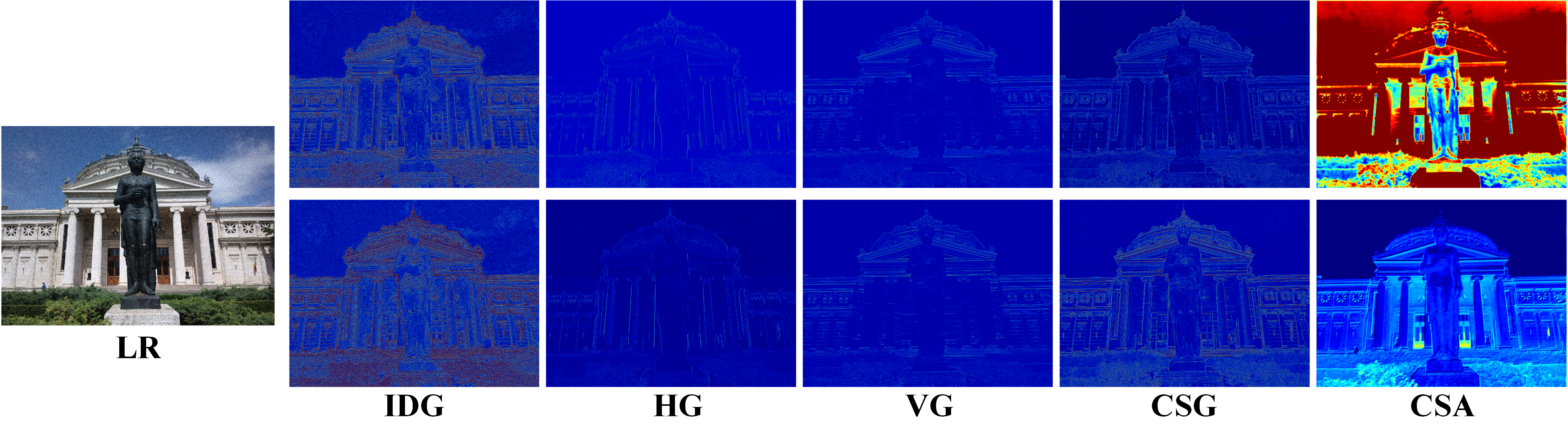}
    \caption{Feature visualization of proposed IDG, HG, VG, CSG, and CSA in our methods.}
    \label{fig:feature1}
\end{figure*}

\begin{table*}[ht]

\centering
\caption{Ablation experiments for our proposed DGConv. Compared with the VConv-based baseline, the performance improvement and degradation are marked as {\textcolor{red}{{\textbf{bold}}$\uparrow$}} and \textbf{\textcolor{blue}{{{{bold$\downarrow$}}}}}, respectively.}

\resizebox{\linewidth}{!}{

\begin{tabular}{cc||ccccccc} 
\hline

Data                                                                                     & Metrics      & VConv            & 5VConvs                                                 & ECB\cite{ECB}                                                   & RepB\cite{ding2021repvgg}                                                  & ACB\cite{ding2019acnet}                                                   & RepSRB\cite{RepSR}                                                & DGConv                            \\ 
\hline
\multirow{2}{*}{Level-I}                                                                 & PSNR         & 26.72        & 26.41                                               & 26.53                                                & 26.56                                                & 26.65                                                & 26.58                                                & 26.77                            \\
                                                                                         & LPIPS        & 0.223         & 0.251                                                & 0.218                                                 & 0.228                                                 & 0.233                                                 & 0.223                                                 & 0.211                             \\
\multirow{2}{*}{Level-II}                                                                & PSNR         & 26.64        & 26.29                                               & 26.50                                                & 26.67                                                & 26.55                                                & 26.51                                                & 26.74                            \\

                                                                                         & LPIPS        & 0.258         & 0.289                                                & 0.252                                                 & 0.257                                                 & 0.262                                                 & 0.260                                                 & 0.243                             \\
\multirow{2}{*}{Level-III}                                                               & PSNR         & 24.17        & 24.01                                               & 24.02                                                & 24.17                                               & 24.22                                                & 24.21                                                & 24.15                            \\
                                                                                         & LPIPS        & 0.402         & 0.445                                                & 0.405                                                 & 0.401                                                 & 0.405                                                 & 0.398                                                 & 0.392                             \\
\multirow{2}{*}{Level-IV}                                                                & PSNR         & 23.81        & 23.68                                               & 23.69                                                & 23.83                                                & 23.86                                                & 23.85                                                & 23.80                            \\
                                                                                         & LPIPS        & 0.427         & 0.467                                                & 0.429                                                 & 0.426                                                 & 0.429                                                 & 0.435                                                 & 0.414                             \\
\multirow{2}{*}{Level-V}                                                                 & PSNR         & 23.63        & 23.48                                               & 23.52                                                & 23.64                                                & 23.66                                              & 23.66                                               & 23.62                            \\
                                                                                         & LPIPS        & 0.447         & 0.484                                                & 0.446                                                 & 0.447                                                 & 0.450                                                 & 0.447                                                 & 0.435                             \\ 
\hline 
\rowcolor[rgb]{0.937,0.937,0.937} {\cellcolor[rgb]{0.937,0.937,0.937}}                   &  PSNR  &  24.996 &  \textcolor{blue}{\textbf{24.77}$\downarrow$} &  \textcolor{blue}{\textbf{24.85$\downarrow$}} &  \textcolor{blue}{\textbf{24.98$\downarrow$}} &  \textcolor{blue}{\textbf{24.99$\downarrow$}} &  \textcolor{blue}{\textbf{24.96$\downarrow$}} &  \textcolor{red}{\textbf{25.06$\uparrow$}}  \\ \rowcolor[rgb]{0.937,0.937,0.937} \multirow{-2}{*}{\cellcolor[HTML]{EFEFEF}{\color[HTML]{000000}Avg}}
 \multirow{-2}{*}{{\cellcolor[rgb]{0.937,0.937,0.937}}} &  LPIPS &  0.352  &  \textcolor{blue}{\textbf{0.387$\downarrow$}} & \textcolor{red}{ \textbf{0.350$\uparrow$}}                       &  \textcolor{blue}{\textbf{0.352$\downarrow$}}  &  \textcolor{blue}{\textbf{0.356$\downarrow$}}  &  \textcolor{blue}{\textbf{0.352$\downarrow$}}  &  \textcolor{red} {\textbf{0.335$\uparrow$}}  \\
\hline

\end{tabular}
}
\label{tab:ablation}
\end{table*}

\subsection{User study}

To further demonstrate the superiority of DGPNet in real-world super-resolution, we conduct a user study on real-world images from real DLSR datasets \cite{cai2019toward} and RealWorld38 \cite{wang2021real}. Specifically, we involve eleven volunteers, which include five experts with backgrounds in image processing and six users without such expertise. We randomly selected 30 scenes, encompassing areas of flat, regular, and irregular textures. Participants are asked to rate each scene individually, and we aggregate the scores from all users. Each image was scored from 0 to 10, with higher scores signifying a greater level of visual satisfaction. As depicted in Fig. \ref{fig:User_study}, our method achieves the highest average score of 7.67 and the lowest variance score of 0.305, compared with the existing methods, demonstrating that our approach has been shown to significantly outperform a range of contemporary techniques in subjective evaluation, offering greater stability and highlighting the superior visual appeal that our method achieves.

\subsection{Feature visualization}
To demonstrate the efficacy of our proposed learnable regular Directional Gradient convolution (IDG), Vertical
Gradient convolution (VG), Horizontal Gradient (HG) convolution, and Center-Surrounding Gradient (CSG) convolution in capturing directional gradients, we present a visual representation of their feature maps in Fig. \ref{fig:feature1}. It is clear from the visualization that the IDG and CSG convolutions are adept at discerning gradients across a multitude of irregular directions. In contrast, the HG and VG convolutions are specifically attuned to gradients in distinct, regular directions, such as horizontal and vertical. Additionally, the Center-Surrounding Aggregation (CSA) convolution effectively captures the statistical properties of the image, including the distribution of brightness.

\subsection{Ablation Study}
To demonstrate the superiority of DGConv in improving network representation capacity without extra computation costs, we compare our DGConv with several different settings and existing settings: VConv, five VConvs in parallel, RepVGG block used in \cite{ding2021repvgg}, asymmetric convolution block used in \cite{ding2019acnet}, RepSR block used in \cite{RepSR}, and ECB block used in \cite{ECB} denoted as VConv, 5VConvs, RepB, ACB, RepSRB, and ECB, respectively. We use the above settings to retrain SRResNet \cite{SRResNet}, and the results are reported in Table. \ref{tab:ablation}. We can observe that 5VConvs show significant performance degradation in both PNSR and LPIPS compared to a single VConv. This may be due to multiple VConv branches having conflicts in network optimization, resulting in performance degradation. RepB, ACB, and RepSRB are all made up of BN and VConv in different receptive fields, and they still cannot efficiently learn the regular and irregular gradation representations, resulting in performance degradation compared to baseline VConv. Although ECB uses edge detection operators to enhance the perception capacity of image structure, it still has problems of being unable to adaptively perceive direction arrangements on different regions and a lack of adaptive fusion of different gradient representations, resulting in limited representation capacity and performance degradation compared to VConv. The superior performance may be attributed to DGConv's capability to capture diverse directional gradient arrangements through its learnable parameters. By employing kernel-aware adaptive fusion, DGConv is able to selectively assimilate varying gradient representations, differentiating between regular and irregular patterns during the training phase. This nuanced approach leads to a comprehensive enhancement in performance when compared to VConv.

In addition, we conduct extensive ablation experiments for each convolution in DGConv, with the results presented in Table. \ref{tab:Ablation_on_ADAPConv}. Our observations reveal that each proposed convolution plays a pivotal role in enhancing the representational capacity of DGConv, thereby validating the rationale and effectiveness of the proposed DGConv. This also prompts us to explore whether incorporating additional directional arrangement convolutions could further augment DGConv's representational capacity. Moreover, we conduct ablation experiments on the intermediate feature channels within DGPNet, with the findings depicted in Table. \ref{tab:Ablation_on_DGPNet}. The results indicate that DGPNet consistently delivers robust performance across varying intermediate feature channels, underscoring the efficacy and robustness of the proposed AIIBlock in DGPNet.

\begin{table}[t]
\centering
\caption{Ablation experiments for each proposed convolution in DGConv reveal that each one plays a crucial role.}
\resizebox{\linewidth}{!}{
\begin{tabular}{cccccc||cc}
\hline
{VConv}                           & {CSG}                         & {HG}                         & {VG}                         & {CAS}      
& {IDG} 
& PSNR$\uparrow$      & LPIPS$\downarrow$   \\ 
\hline
                                                \ding{55} & \ding{51}  & \ding{51}  & \ding{51}  & \ding{51}  & \ding{51}  & 25.38 & 0.321 \\
\ding{51}  &  \ding{55}                                                & \ding{51}  & \ding{51}  & \ding{51}  & \ding{51}  & 25.30 & 0.323 \\
\ding{51}  & \ding{51}  &   \ding{55}                                               & \ding{51}  & \ding{51}  & \ding{51}  & 25.31 & 0.327 \\
\ding{51}  & \ding{51}  & \ding{51}  &   \ding{55}                                               & \ding{51}  & \ding{51}  & 25.29 & 0.328 \\
\ding{51}  & \ding{51}  & \ding{51}  & \ding{51}  &  \ding{55}                        & \ding{51}  & 25.28 & 0.325 \\
\ding{51}  & \ding{51}  & \ding{51}  & \ding{51}  & \ding{51}  & \ding{55}  & 25.21 & 0.328 \\ \rowcolor[HTML]{EFEFEF} 
\ding{51}  & \ding{51}  & \ding{51}  & \ding{51}  & \ding{51}  & \ding{51}  & \textbf{25.41} & \textbf{0.309} \\ \hline 

\end{tabular}
}
\label{tab:Ablation_on_ADAPConv}
\end{table}
\begin{table}[t]
\centering
\caption{Ablation experiments on the number of intermediate feature channels in DGPNet. Note that the feature channels for DGPNet\_tiny and DGPNet are 32 and 64, respectively. It demonstrates that, with different channel counts, DGPNet still achieves commendable performance and efficiency.}
\resizebox{\columnwidth}{!}{
\begin{tabular}{c||cccc}
\hline
            Feature Channel                   & PSNR$\uparrow$     & LPIPS$\downarrow$   & FLOPs(G)    & Params(M) \\ \hline \hline
32          & 25.37 & 0.335 & 90.419   & 1.123  \\
48            & 25.40 & 0.317 & 202.670 & 2.522 \\
64 & {25.41} & {0.309} & {359.615}  & {4.480} \\ \hline
\end{tabular}}
\label{tab:Ablation_on_DGPNet}
\end{table}
\begin{table}[t]
\centering
\caption{Ablation experiment on CSA and CSG convolutions. DGConv* represents using the central value. This validates the effectiveness of using the local mean in CSA and CSG.}
\resizebox{\columnwidth}{!}{
\begin{tabular}{ccccc}
\hline
\hline
                           & \multicolumn{2}{c}{VDSR}                         & \multicolumn{2}{c}{EDSR}    \\

                           
\multirow{-2}{*}{Baseline} & DGConv* & DGConv                                  & DGConv* & DGConv                                  \\ \hline
PSNR$\uparrow$                       & 24.02   & \cellcolor[HTML]{EFEFEF}\textbf{24.147} & 24.947  & \cellcolor[HTML]{EFEFEF}\textbf{25.186} \\
LPIPS$\downarrow$                      & 0.413   & \cellcolor[HTML]{EFEFEF}\textbf{0.408}  & 0.352   & \cellcolor[HTML]{EFEFEF}\textbf{0.343}  \\ \hline \hline
\end{tabular}
}
\label{tab:localmean}
\end{table}
\begin{table}
\centering
\caption{The performance and computational cost of VConv-based and DGConv-based baseline. VConv-based and DGConv-based methods are marked as bold and \textbf{bold*}. DGConv-based methods exhibit \textbf{superior performance} while \textbf{maintaining consistent computational costs}, compared to VConv.}
\resizebox{\linewidth}{!}{
\begin{tabular}{ccccc}
\hline \hline
\multicolumn{1}{c}{Methods} & PSNR$\uparrow$            & LPIPS$\downarrow$          & FLOPs(G)        & Params(M)        \\ \hline
SRResNet                     & 24.996          & 0.352          & 121.928          & 1.554          \\
\rowcolor[HTML]{EFEFEF} 
\textbf{SRResNet*}           & \textbf{25.064} & \textbf{0.335} & \textbf{121.928} & \textbf{1.554} \\ \hline
EDSR                         & 24.995          & 0.353          & 129.969          & 1.518          \\
\rowcolor[HTML]{EFEFEF} 
\textbf{EDSR*}               & \textbf{25.186} & \textbf{0.343} & \textbf{129.969} & \textbf{1.518} \\ \hline
VDSR                         & 23.651          & 0.434          & 699.409          & 0.667          \\
\rowcolor[HTML]{EFEFEF} 
\textbf{VDSR*}               & \textbf{24.147} & \textbf{0.408} & \textbf{699.409} & \textbf{0.667}          \\ \hline \hline
\end{tabular}}

\label{tab:cost_free}
\end{table}

\subsection{Differences from Traditional Handcrafted Operators}

Traditional static handcrafted operators, such as the Laplacian and Sobel, employ fixed kernels to extract image features. The intuitive idea is to use them as a preprocessing step in CNNs to extract gradient features. However, this approach encounters two issues: (1) It lacks the ability to adaptively perceive gradients in different directional arrangements, resulting in limited representational capacity. To further validate this, we conduct an ablation study, replacing our learnable convolution with a fixed handcrafted convolution in the training model. We observe that after employing traditional handcrafted kernels, the performance of DGPNet and SRResNet deteriorated by 0.53 dB and 0.87 dB in PSNR, respectively, demonstrating the superiority of our proposed learnable convolution. (2) Incorporating a preprocessing step in CNNs inevitably introduces additional computational overhead. For instance, for a $3 \times 3$ Laplacian operator and a $3 \times 3$ VConv kernel, the preprocessing method nearly doubles the parameters and FLOPs of the model, significantly limiting the practicality of this approach. In contrast, our proposed method not only adaptively captures gradient information in different directions but also incurs only half the computational cost of the preprocessing approach.

\subsection{Rationale for using local statistical mean in CSG and CSA}
Utilizing the central value directly in our learnable center-surrounding gradient may overlook the crucial center-directional gradient and struggle to capture all noise when the central and surrounding noise values are similar. To tackle this issue, we propose employing the local mean as a reference for the gradient to capture a variety of directional gradients. This approach can mitigate these challenges and provide nine-direction gradient arrangement information extending from the periphery to the center for representation learning. However, for center-surrounding aggregation convolutions, our objective is to capture local contrast statistics (\emph{i.e.}, low-frequency statistical properties). Relying on the value at the central position as the statistical property may lead to inaccuracies. Hence, we propose using the mean component of the local statistical features directly to circumvent these errors. In this regard, we incorporate the local mean to enhance the proportion of statistical information and propose a learnable center-surrounding aggregation convolution to recover the statistical properties of images.

To validate the efficacy of employing the local mean in DGConv, we conduct ablation experiments comparing the use of the local mean and the central value in our convolutions. We create DGConv* by replacing the local mean with the central value and testing it on VDSR and EDSR baselines. The results are presented in Table. \ref{tab:localmean}, which indicates that the performance of DGConv declined after substituting the local mean with the central value. This confirms the rationality and superiority of utilizing the local mean in DGConv.

\subsection{The evidence of cost-free improvement of DGConv}
\label{The evidence of cost-free improvement of DGConv}

To further validate that our proposed DGConv can improve performance without introducing additional computational cost, we select three baselines (SRResNet, EDSR, and VDSR), train them using VConv and DGConv, and compute their parameter and FLOPs. The results are shown in Table. \ref{tab:cost_free}. We can observe that compared to the VConv-based model, the DGConv-based model is able to improve both PSNR and LPIPS metrics, demonstrating the effectiveness of DGConv. Furthermore, we can observe that the computational overhead cost (Params and FLOPs) of DGConv-based networks is the same as that of VConv-based networks. This demonstrates the \textbf{superiority} and \textbf{plug-and-play} properties of DGConv.

\section{Conclusion}
In this paper, we propose a novel plug-and-play DGConv to facilitate the extraction and balance of detail-relevant, contrast-relevant, and degradation-relevant features by embedding directional gradient operation into the convolutional process. With the proposed equivalent parameter fusion, the computational complexity of DGConv is the same as that of vanilla convolution during inferencing. To adeptly balance the enhancement of texture and contrast while meticulously investigating the interdependencies between them, we propose a simple yet efficient AIIBlock and DGPNet that achieve state-of-the-art performance on various datasets with low computational complexity.

In the future, we will consider proposing more novel convolutions with the perception capacity of complete gradient arrangements to cover more texture and degradation cues in real-world scenarios. At the same time, we will consider validating and extending the proposed DGConv to more image and video super-resolution tasks and image restoration tasks.

{
\bibliographystyle{IEEEtran}
\bibliography{IEEEabrv,egbib}
}

\begin{IEEEbiography}[{\includegraphics[width=1in,height=1.25in,clip,keepaspectratio]{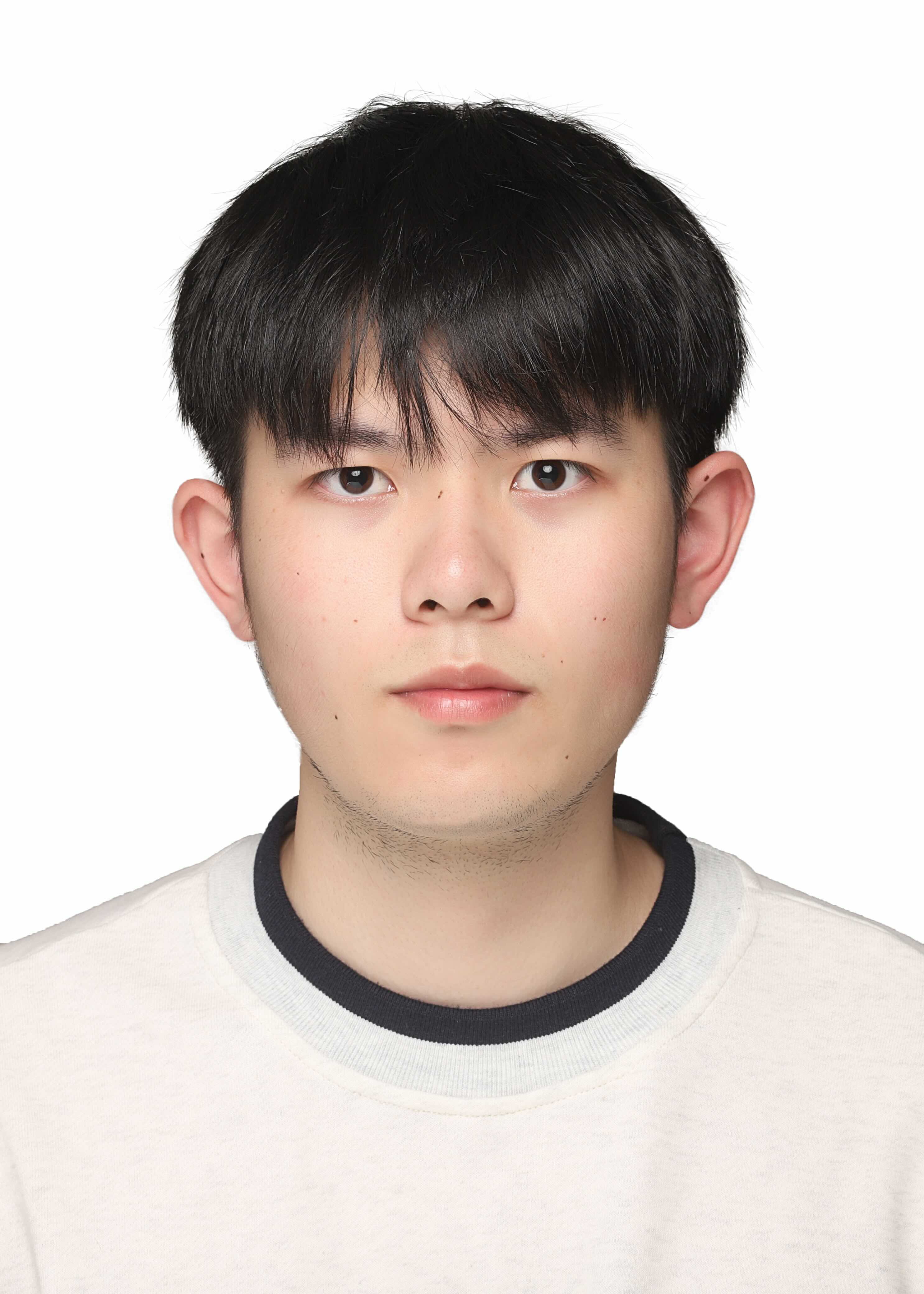}}]
{Long Peng} is pursuing a master's degree at the University of Science and Technology of China. He has published papers in the IEEE International Conference on Computer Vision and Pattern Recognition (CVPR), IEEE Transactions on Multimedia (TMM), IEEE Transactions on Artificial Intelligence (TAI), and IEEE Signal Processing Letters (SPL), among others.  His research interests are image processing, low-level vision, and computer vision.
\end{IEEEbiography}

\begin{IEEEbiography}[{\includegraphics[width=1in,height=1.25in,clip,keepaspectratio]{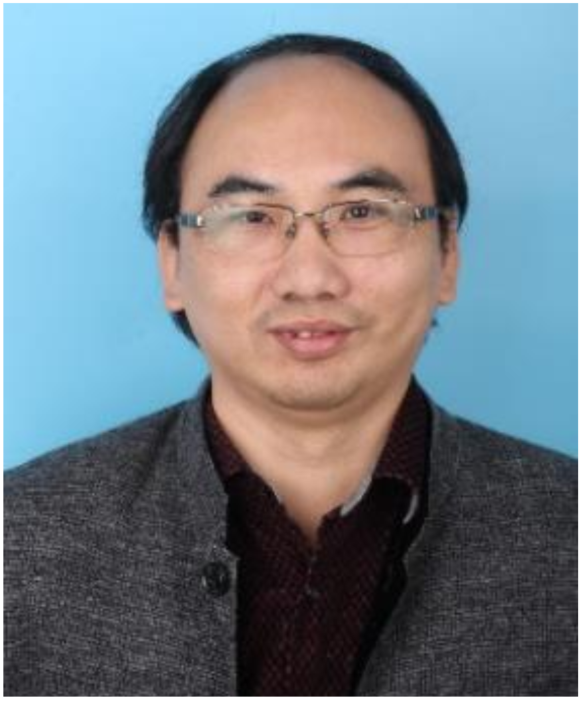}}]
{Yang Cao} (Member, IEEE) was born in 1980.
He received the B.S. and Ph.D. degree in information engineering from Northeastern University, Shenyang, China, in 1999 and 2004, respectively. Since 2004, he has been with the Department of Automation, University of Science and Technology of China, Hefei, China, where he is currently an Associate Professor. His current
research interests include machine learning and computer vision. Dr. Cao is a member of the IEEE Signal Processing Society.
\end{IEEEbiography}

\begin{IEEEbiography}[{\includegraphics[width=1in,height=1.25in,clip,keepaspectratio]{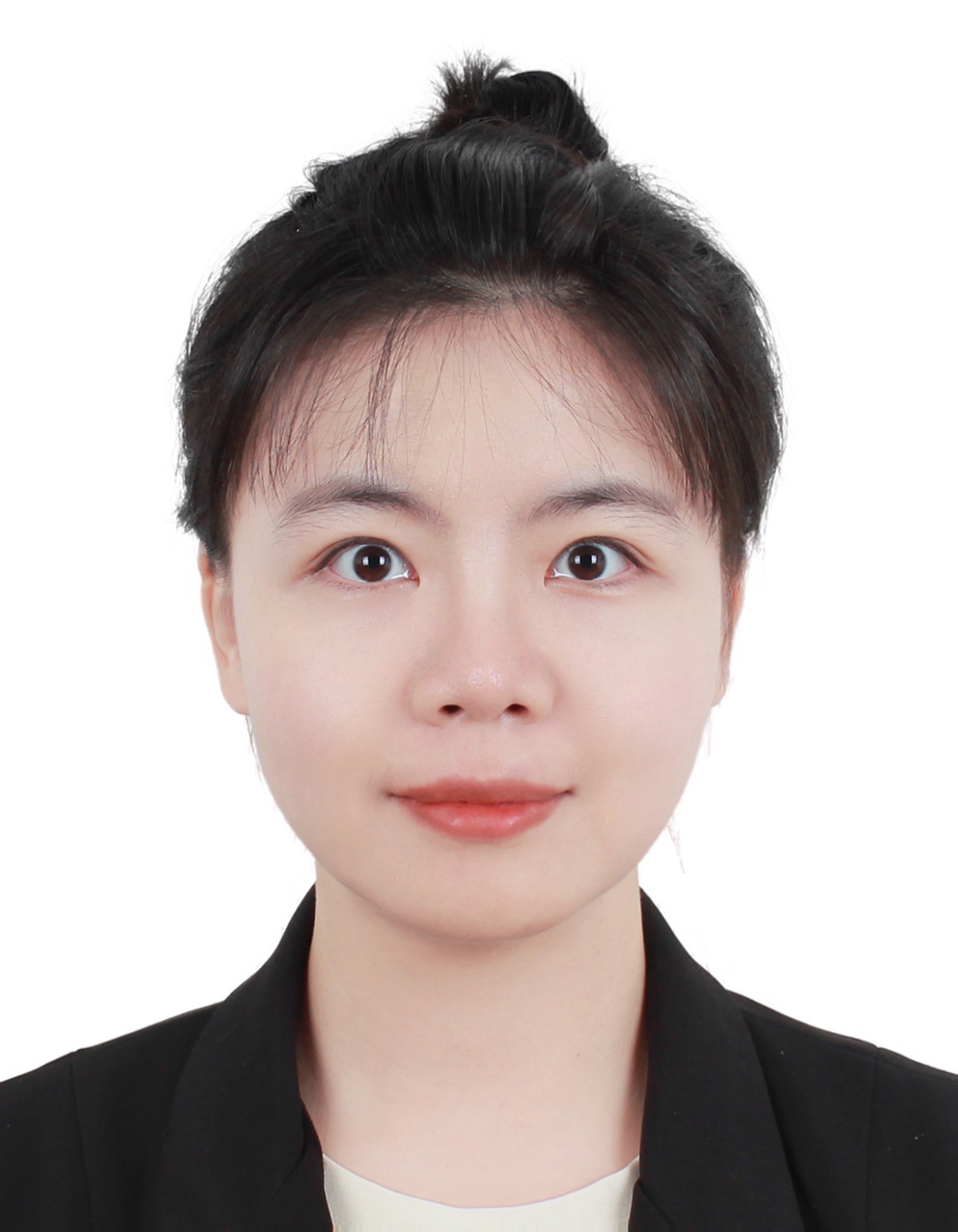}}]
{Ren-Jing Pei} obtained her B.E. degree from Wuhan University and her Ph.D. from the Institute of Automation, Chinese Academy of Sciences, Beijing. Currently, she works as a researcher at Noah’s Ark Lab, Huawei Technologies Ltd. Her research interests encompass low-level vision, multimodal understanding, and generation.
\end{IEEEbiography}

\begin{IEEEbiography}[{\includegraphics[width=1in,height=1.25in,clip,keepaspectratio]{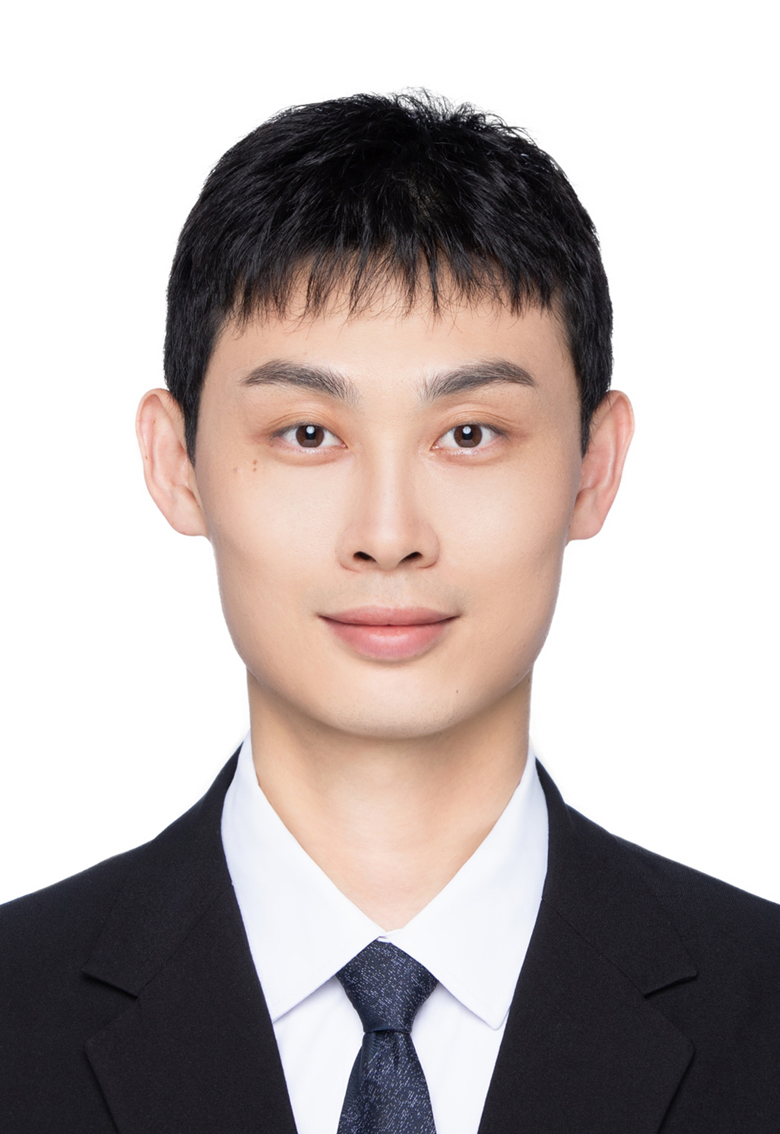}}]
{Wenbo Li} received the BEng and MS degrees from Shanghai Jiao Tong University, in 2016 and 2019, and the PhD degree in the Department of Computer Science and Engineering from The Chinese University of Hong Kong, in 2023. His primary research interests lie in low-level computer vision and AIGC. His paper was selected in the CVPR 2022 Best Paper Finalists. He also serves as a reviewer for IEEE Transactions on Pattern Analysis and Machine Intelligence, International Journal of Computer Vision, IEEE Transactions on Image Processing, ICML, ICLR, NeurIPS, CVPR, etc.
\end{IEEEbiography}

\begin{IEEEbiography}[{\includegraphics[width=1in,height=1.25in,clip,keepaspectratio]{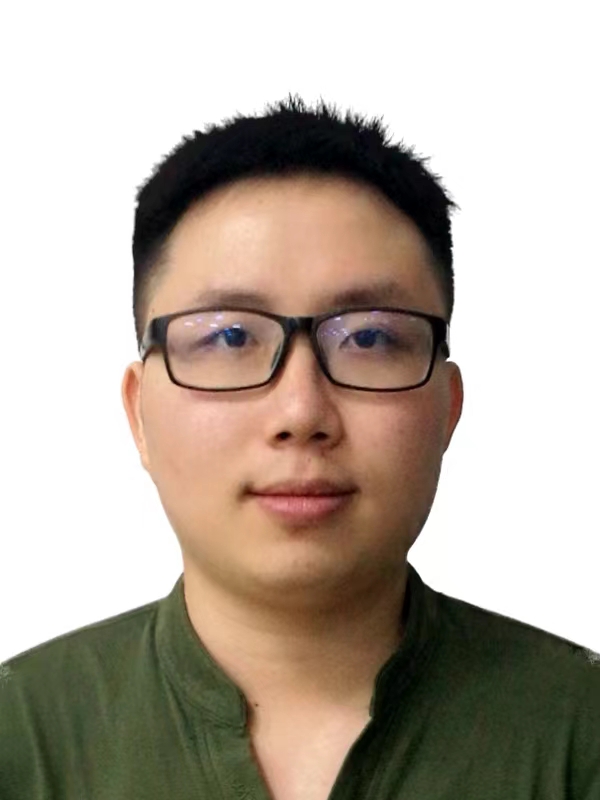}}]
{Jiaming Guo} received the M.E. degree from Sun Yat-Sen university, Guangzhou, China, in 2019. He is currently a researcher with Noah's Ark Lab, Central Research Institute, 2012 Laboratories, Huawei Technologies Co., Ltd. His research interests include super-resolution, image restoration, efficient model etc. He has published several papers in international journals and conference proceedings, e.g. CVPR and ECCV.
\end{IEEEbiography}

\begin{IEEEbiography}[{\includegraphics[width=1in,height=1.25in,clip,keepaspectratio]{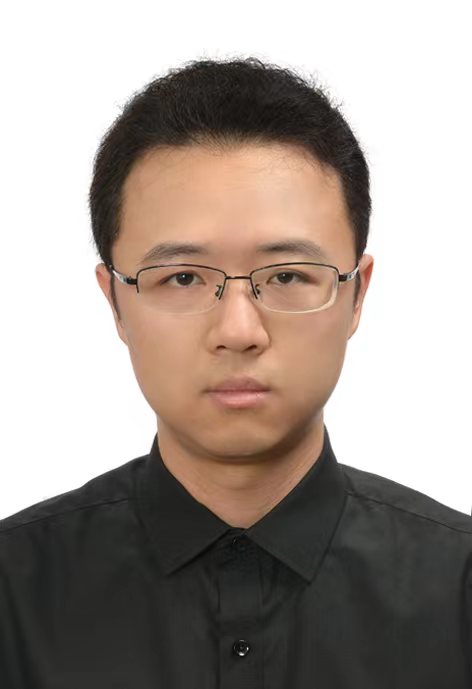}}]
{Xueyang Fu} (Member, IEEE) received the Ph.D. degree in signal and information processing from Xiamen University, Xiamen, China, in 2018. He was a Visiting Scholar at Columbia University, New York, NY, USA, sponsored by the China Scholarship Council, from 2016 to 2017. He is currently an Associate Researcher with the Department of Automation, University of Science and Technology of China, Hefei, China. His research interests include machine learning and image processing.
\end{IEEEbiography}

\begin{IEEEbiography}[{\includegraphics[width=1in,height=1.25in,clip,keepaspectratio]{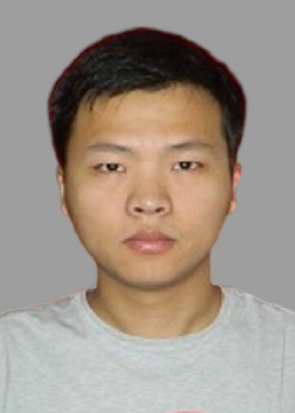}}]
{Yang Wang} (Member, IEEE) received the Ph.D. degree in control science and engineering from the University of Science and Technology of China in 2021. He is currently an associate professor at the School of Information Engineering at Chang’an University, Xi’an, China. His research interests include
machine learning and image processing
\end{IEEEbiography}

\begin{IEEEbiography}[{\includegraphics[width=1in,height=1.25in,clip,keepaspectratio]{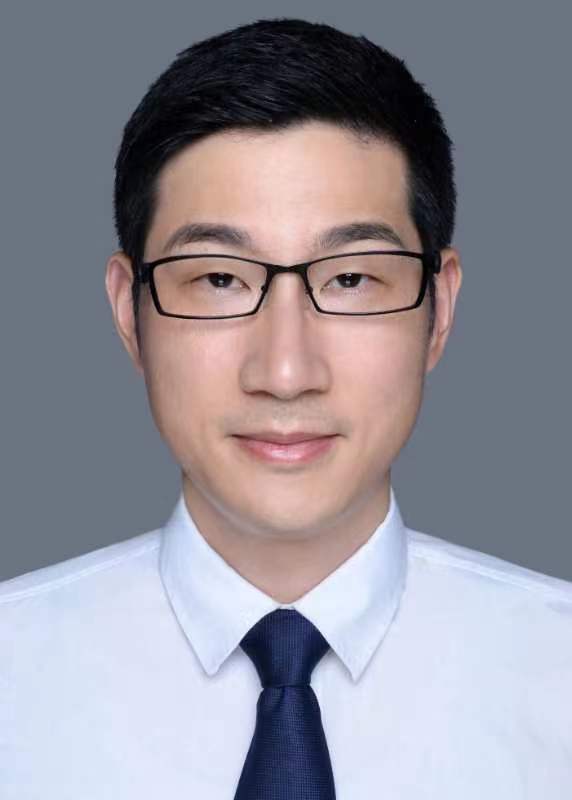}}]
{Zheng-Jun Zha} (Member, IEEE) received the B.E. and Ph.D. degrees from the University of Science and Technology of China, Hefei, China, in 2004 and 2009, respectively.

He is currently a Full Professor with the School of Information Science and Technology and the Executive Director of the National Engineering Lab- oratory for Brain-Inspired Intelligence Technology and Application (NEL-BITA), University of Science and Technology of China. He has authored or co-authored more than 200 papers in his research field with a series of publications in top journals and conferences, which include multimedia analysis and understanding, computer vision, pattern recognition, and brain-inspired intelligence

Dr. Zha was a recipient of multiple paper awards from prestigious conferences, including the Best Paper/Student Paper Award at the Association for Computing Machinery (ACM) Multimedia and AAAI Distinguished Paper. He serves/served as an Associate Editor for IEEE TRANSACTIONS ON MULTIMEDIA, IEEE TRANSACTIONS ON CIRCUITS AND SYSTEMS FOR VIDEO TECHNOLOGY, and ACM Transactions on Multimedia Computing, Communications, and Applications.
\end{IEEEbiography}

\end{document}